\pdfoutput=1
\documentclass[11pt,twoside,a4paper,cmspaper,final,collab]{cms-tdr}
\def\svnVersion{c4de6b1}\def\svnDate{2026/10/23}\def\cmsCernNoTag{CERN-EP-2026-111}\def\cmsCernDate{\today}\def\cmsMessage{Published in the Journal of High Energy Physics as \href{http://dx.doi.org/10.1007/JHEP10(2026)167}{\doi{10.1007/JHEP10(2026)167}.}}
\begin{document}\cmsNoteHeader{JME-25-001}

\newcommand{\PY}{{\HepParticle{Y}{}{}}\xspace}
\newcommand{\PG}{{\HepParticle{G}{}{}}\xspace}
\newcommand{\PS}{{\HepParticle{S}{}{}}\xspace}
\providecommand{\PWst}{\HepParticle{\PW}{}{\ast}\xspace}

\newcommand{\kapl}{\ensuremath{\kappa_{\lambda}}\xspace}
\newcommand{\kapv}{\ensuremath{\kappa_{\PV}}\xspace}
\newcommand{\kapvv}{\ensuremath{\kappa_{2\PV}}\xspace}
\newcommand{\HY}{\ensuremath{{\PH\PY}}\xspace}
\newcommand{\HH}{\ensuremath{{\PH\PH}}\xspace}
\newcommand{\HHVV}{\ensuremath{{\PH\PH\PV\PV}}\xspace}
\newcommand{\bbbb}{\ensuremath{\bbbar\bbbar}\xspace}
\newcommand{\mSD}{\ensuremath{m_\text{SD}}\xspace}
\newcommand{\ssbar}{\PQs{}\PAQs\xspace}

\newcommand{\hbb}{\ensuremath{\PH\to\bbbar}\xspace}
\newcommand{\hcc}{\ensuremath{\PH\to\ccbar}\xspace}
\newcommand{\hssq}{\ensuremath{\PH\to\ssbar}\xspace}
\newcommand{\hqq}{\ensuremath{\PH\to\qqbar}\xspace}
\newcommand{\htata}{\ensuremath{\PH\to\PGtp\PGtm}\xspace}
\newcommand{\qqq}{\ensuremath{3\PQq}\xspace}
\newcommand{\qqqq}{\ensuremath{4\Pq}\xspace}
\newcommand{\ww}{\ensuremath{\PW\PW}\xspace}
\newcommand{\bbww}{\ensuremath{\bbbar\PW\PW}\xspace}
\newcommand{\hww}{\ensuremath{\PH\to\ww}\xspace}
\newcommand{\hwws}{\ensuremath{\PH\to\PW\PWst}\xspace}
\newcommand{\wws}{\ensuremath{\PW\PWst}\xspace}
\newcommand{\yww}{\ensuremath{\PY\to\ww}\xspace}
\newcommand{\tbqq}{\ensuremath{\PQt\to\PQb\PQq\PAQq}\xspace}
\newcommand{\hwwq}{\ensuremath{\PH\to\PW\PWst\to\qqqq}\xspace}
\newcommand{\hwwqnoast}{\ensuremath{\PH\to\ww\to\qqqq}\xspace}
\newcommand{\HHbbVV}{\ensuremath{{\PH\PH\to\bbbar\PW\PWst}}\xspace}

\newcommand{\THWW}{\ensuremath{T_{\PH\PW\PW}}\xspace}
\newcommand{\THWWnotop}{\ensuremath{T_{\PH\PW\PW}^\text{No top}}\xspace}
\newcommand{\PQCD}{\ensuremath{P_\text{QCD}}\xspace}
\newcommand{\PTop}{\ensuremath{P_\text{top}}\xspace}
\newcommand{\PHWWqqqq}{\ensuremath{P_{\PH\PW\PW4\PQq}}\xspace}
\newcommand{\PHWWqqq}{\ensuremath{P_{\PH\PW\PW3\PQq}}\xspace}
\newcommand{\PWqq}{\ensuremath{P_{\PW\PQq\PQq}}\xspace}
\newcommand{\PHqqqq}{\ensuremath{P_{\PH4\PQq}}\xspace}

\newcommand{\tbw}{\ensuremath{\PQt\to\PQb\PW}\xspace}
\newcommand{\tae}{\ensuremath{\PGt_{\Pe}}\xspace}
\newcommand{\tam}{\ensuremath{\PGt_{\PGm}}\xspace}

\newcommand{\mpeak}{\ensuremath{{m_\text{peak}}}\xspace}
\newcommand{\mg}{\ensuremath{{m_\PG}}\xspace}
\newcommand{\mh}{\ensuremath{{m_\PH}}\xspace}
\newcommand{\my}{\ensuremath{{m_\PY}}\xspace}
\newcommand{\mt}{\ensuremath{{m_\PQt}}\xspace}
\newcommand{\mw}{\ensuremath{{m_\PW}}\xspace}
\newcommand{\ms}{\ensuremath{{m_\PS}}\xspace}

\newcommand{\msd}{\ensuremath{{m_\mathrm{SD}}}\xspace}
\newcommand{\mreco}{\ensuremath{{m_\text{reco}}}\xspace}
\newcommand{\mreg}{\ensuremath{{m_\text{reg}}}\xspace}
\newcommand{\epsb}{\ensuremath{{\epsilon_\mathrm{B}}}\xspace}

\newcommand{\JHUGEN}{{\textsc{JHUGen}}\xspace}

\newcommand{\partmodel}{\textsc{ParT}\xspace}
\newcommand{\particlenet}{\textsc{ParticleNet}\xspace}
\newcommand{\particlenetmd}{\textsc{ParticleNet-MD}\xspace}
\newcommand{\deepak}{\textsc{DeepAK8-MD}\xspace}

\newcommand{\ifrac}[2]{\ensuremath{{#1}/{#2}}\xspace}
\newcommand{\hhexp}{69\xspace}
\newcommand{\taggername}{\textsc{ParT}\xspace}
\newlength\cmsTabSkip\setlength{\cmsTabSkip}{2ex}

\cmsNoteHeader{JME-25-001}
\title{Particle transformers for identifying Lorentz-boosted Higgs bosons decaying to a pair of \texorpdfstring{\PW}{W} bosons}

\date{\today}

\abstract{
    A novel deep neural network classifier, a ``particle transformer'' (\taggername), is introduced for the identification of highly Lorentz-boosted resonances reconstructed as single, multipronged jets in measurements and searches performed by the CMS Collaboration at the CERN LHC.  Based on a self-attention mechanism that allows the model to weigh the importance of different particles, \taggername is trained on a wide variety of topologies, notably demonstrating strong performance for the first time on jets originating from boosted Higgs boson decays to \PW bosons.  The \taggername algorithm achieves a tagging efficiency of more than 50\% for such jets at a background efficiency of 1\%, while maintaining decorrelation from the jet mass.  A calibration is performed in proton-proton collision data collected by CMS at a center-of-mass energy of 13\TeV, with a data set corresponding to a total luminosity of 138\fbinv.  Data-to-simulation selection efficiency scale factors are measured to be in the 0.9--1.0 range, with relative uncertainties between 7 and 23\%.  The tagging capability of \taggername enhances the sensitivity of standard model measurements and searches for beyond-the-standard-model resonances decaying to hadronic diboson systems.
}

\hypersetup{
    pdfauthor={CMS Collaboration},
    pdftitle={Particle transformers for identifying Lorentz-boosted Higgs bosons decaying to a pair of W bosons},
    pdfsubject={CMS},
    pdfkeywords={CMS, HWW, ML, jets}}

\maketitle

\section{Introduction}
\label{sec:intro}

Since discovering the Higgs boson (\PH)~\cite{Higgs-Discovery_CMS, Higgs-Discovery_CMS_long, Higgs-Discovery_ATLAS}, the CMS and ATLAS Collaborations at CERN have been focused on precisely measuring its properties and interactions.
Identifying its hadronic final states is an important avenue for doing so, as well as for searching for new physics processes.
Traditionally, however, they have been considered challenging at a hadron collider because of the large background from standard model (SM) events composed uniquely of jets produced through the strong interaction, referred to as quantum chromodynamics (QCD) multijet events, as well as resonant processes such as top quark pair (\ttbar) production.

    {
        \tolerance=1000
        Over the past few years, the advent of progressively more advanced deep-learning (DL) techniques for jet identification, combined with the exploration of the highly Lorentz-boosted regime, has allowed the LHC experiments to steadily address this challenge.
        By exploiting the detailed low-level features of jets, including the kinematic, identification, and trajectory-related information of particle constituents and the presence and characteristics of secondary vertices, several DL algorithms have significantly improved our performance in tagging jets originating from decays of boosted resonances.
        These include \textsc{DeepAK8}~\cite{CMS:2020poo} and \textsc{DeepDoubleX}~\cite{CMS-DP-2022-041}, which combine convolutional and recurrent neural networks; \textsc{Xbb}~\cite{ATL-PHYS-PUB-2020-019}, which combines recurrent and feedforward neural networks; \textsc{DeXTer}~\cite{ATL-PHYS-PUB-2022-042}, a deep-set-based neural network; and graph neural network (GNN) architectures, such as interaction networks~\cite{Moreno:2019bmu, Moreno:2019neq}, \particlenet~\cite{Qu:2019gqs, CMS-DP-2020-002, CMS-DP-2021-017}, and \textsc{GN2X}~\cite{ATL-PHYS-PUB-2023-021}.
        At a typical working point with a signal efficiency of about 50\%, these advanced algorithms have improved background rejection by nearly an order of magnitude compared with simpler algorithms based on high-level jet features.
        They have been successfully employed to set some of the most stringent constraints on SM \PH to bottom quark (\hbb) decays in the gluon fusion (ggF) and vector boson fusion (VBF) production modes~\cite{CMS:2024ddc}, the charm Yukawa coupling~\cite{CMS:2022psv}, \PH pair (\HH) production, and the quartic coupling of two Higgs bosons and two weak (\PV) vector bosons \HHVV~\cite{CMS:2022gjd,ATLAS-CONF-2024-003}, as well as many beyond the SM (BSM) theories~\cite{CMS:2022suh, CMS:2024pjq, CMS:2021roc, CMS:2021mjl, CMS:2022lqh, ATLAS:2023azi}.
        Beyond the LHC experiments, phenomenological studies with open data sets have studied a broad range of DL techniques for jet tagging, including image-based convolutional neural networks~\cite{deOliveira:2015xxd}, set-based energy-flow networks~\cite{Komiske:2018cqr}, theory-inspired architectures~\cite{Dreyer:2020brq}, attention-based GNNs~\cite{Mikuni:2020wpr}, Lorentz-equivariant neural networks~\cite{pmlr-v119-bogatskiy20a,Bogatskiy:2023nnw,Brehmer:2024yqw,NEURIPS2024_277628cf}, and transformer-based architectures~\cite{Qu:2022mxj}.
        \par
    }

These previous DL models have primarily focused on identifying \emph{two-pronged}~\cite{Larkoski:2017jix} large-radius jets, originating from the hadronization of two \PQb or two \PQc quarks produced in \hbb and \hcc decays.
In this paper, we introduce a new model based on the particle transformer (\taggername) architecture~\cite{Qu:2022mxj}, and generalize it to identify large-radius jets with three or more prongs as well.
These include the hadronic decay of $\PQt\to\PW\PQb\to\Pq\Pq\PQb$, and of the \PH to a \PW boson pair and then to four quarks, \hwwq.
Throughout this paper, \hwws denotes the SM decay, in which one of the two \PW bosons is off shell, while \hww is used in the more general case where both \PW bosons may be either on or off shell, as for the Higgs boson-like scalars of varying mass used to train the algorithm (Section~\ref{sec:event_sim}).
While three- and four-pronged jet taggers have been explored in phenomenological studies~\cite{Kasieczka:2019dbj,Bradshaw:2019ipy,Aguilar-Saavedra:2020sxp,Aguilar-Saavedra:2020uhm,Aguilar-Saavedra:2023pde}, the \hwwqnoast topology in particular has seldom been probed experimentally at the LHC and only in the context of BSM searches~\cite{CMS:2015fdn}.
Since the \hwwq decay has the second largest branching fraction after \hbb, this presents an opportunity to significantly constrain SM Higgs boson parameters, such as the quartic coupling between the Higgs and \PV boson, \HHVV.
It may also facilitate searches for a variety of new physics signatures, such as heavy Higgs bosons and two-Higgs-doublet models~\cite{Branco:2011iw, Craig:2013hca, Domingo:2022kfm}, where a large fraction of the signal exhibits a signature of a boosted \PW boson pair decaying hadronically, as well as BSM resonances decaying to boosted multiple vector boson systems~\cite{Agashe:2016kfr, Agashe:2017wss, Ren:2014bza, Kuang:2015gka}.

Significant challenges to \hwws tagging include the need for stable performance across a wide range of jet masses (\ie, \emph{mass decorrelation}), the asymmetric kinematic properties of the intermediate vector bosons, and the lack of a similar and abundant SM proxy for its calibration.
The requirement of mass decorrelation enables searching for resonances in the jet mass distribution, while constraining backgrounds in data sidebands~\cite{Dolen:2016kst}.
We address these challenges with a novel data generation and training scheme, and the use of a new calibration technique based on the primary Lund jet plane (LJP)~\cite{Dreyer:2018nbf, CMS:2025eyd}.
The calibration is performed in data collected in 2016--2018 with the CMS detector at the LHC in proton-proton ($\Pp\Pp$) collisions at a center-of-mass energy of 13\TeV, with a data set corresponding to an integrated luminosity of 138\fbinv.
Finally, we discuss an application of these techniques to the first search for SM \HH production in the all-hadronic \bbww decay mode~\cite{CMS:2025ngq}.

The rest of this paper is organized as follows.
Section~\ref{sec:cms_detector} describes the CMS detector, Section~\ref{sec:event_reco} explains the relevant event reconstruction algorithms, and Section~\ref{sec:event_sim} details the event simulations used in this work.
Section~\ref{sec:tagger} then introduces the \taggername model architecture, input features, and training, and Section~\ref{sec:part_perf} presents its performance and calibration for \hwwq tagging compared with a previous state-of-the-art algorithm.
Section~\ref{sec:hhbbvv} discusses its impact on the all-hadronic \HHbbVV search.
Finally, a summary is presented in Section~\ref{sec:conclusion}.

\section{The CMS detector}
\label{sec:cms_detector}

The CMS apparatus~\cite{CMS:2008xjf,CMS:2023gfb} is a multipurpose, nearly hermetic detector, designed to trigger on~\cite{CMS:2020cmk,CMS:2016ngn,CMS:2024aqx} and identify electrons, muons, photons, and (charged and neutral) hadrons~\cite{CMS:2020uim,CMS:2018rym,CMS:2014pgm}.
Its central feature is a superconducting solenoid of 6\unit{m} internal diameter, providing a magnetic field of 3.8\unit{T}.
Within the solenoid volume are a silicon pixel and strip tracker, a lead tungstate crystal electromagnetic calorimeter (ECAL), and a brass and scintillator hadron calorimeter (HCAL), each composed of a barrel and two endcap sections.
Forward calorimeters extend the pseudorapidity ($\eta$) coverage provided by the barrel and endcap detectors.
Muons are reconstructed using gas-ionization detectors embedded in the steel flux-return yoke outside the solenoid.
More detailed descriptions of the CMS detector, together with a definition of the coordinate system used and the relevant kinematic variables, can be found in Refs.~\cite{CMS:2008xjf,CMS:2023gfb}.

Events of interest are selected using a two-tiered trigger system.
The first level, composed of custom hardware processors, uses information from the calorimeters and muon detectors to select events at a rate of around 100\unit{kHz} within a fixed latency of about 4\mus~\cite{CMS:2020cmk}.
The second level, known as the high-level trigger, consists of a farm of processors running a version of the full event reconstruction software optimized for fast processing, and reduces the event rate to a few \unit{kHz} before data storage~\cite{CMS:2016ngn}.

The silicon tracker used in 2016 measured charged particles within the range $\abs{\eta} < 2.5$.
For nonisolated particles of $1 < \pt < 10\GeV$ and $\abs{\eta} < 1.4$, the track resolutions were typically 1.5\% in \pt and 25--90 (45--150)\mum in the transverse (longitudinal) impact parameter~\cite{CMS:2014pgm}.
At the start of 2017, a new pixel detector was installed~\cite{Phase1Pixel}; the upgraded tracker measured particles up to $\abs{\eta} = 3.0$ with typical resolutions of 1.5\% in \pt and 20--75\mum in the transverse impact parameter~\cite{DP-2020-049} for nonisolated particles of $1 < \pt < 10\GeV$.
According to simulation studies~\cite{DP-2017-015}, similar improvement is expected in the longitudinal direction.

\section{Event reconstruction}
\label{sec:event_reco}

The main physics objects of interest in this paper are large-radius jets that originate from the hadronic decays of boosted resonances and are reconstructed from the final-state particles produced in the $\Pp\Pp$ collisions.
Several different algorithms are used to reconstruct these jets, estimate their properties (such as their mass and momentum), identify their origin, and correct them for detector effects.
This section briefly describes the relevant algorithms used in this paper.

A particle-flow (PF) algorithm~\cite{CMS:2017yfk} aims to reconstruct and identify each individual particle in an event, with an optimized combination of information from the various elements of the CMS detector.
In this process, particles are identified exclusively as charged or neutral hadrons, photons, electrons, or muons, and are collectively referred to as PF candidates.
The energy of photons is obtained from the ECAL measurement.
The energy of electrons is determined from a combination of the electron momentum at the primary interaction vertex as determined by the tracker, the energy of the corresponding ECAL cluster, and the energy sum of all bremsstrahlung photons spatially compatible with originating from the electron track.
The energy of muons is obtained from the curvature of the corresponding track.
The energy of charged hadrons is determined from a combination of their momentum measured in the tracker and the matching ECAL and HCAL energy deposits, corrected for the response function of the calorimeters to hadronic showers.
Finally, the energy of neutral hadrons is obtained from the corresponding corrected ECAL and HCAL energies.
The primary vertex is taken to be the vertex corresponding to the hardest scattering in the event, evaluated using tracking information alone, as described in Section 9.4.1 of Ref.~\cite{CMS-TDR-15-02}.
Secondary vertices (SVs), such as those found in \PQb hadron decays, are reconstructed using the inclusive vertex finder algorithm, which identifies tracks with large impact parameters~\cite{Tsai:2025yrl}.

For each event, hadronic jets are clustered from these reconstructed particles using the infrared- and collinear-safe anti-\kt algorithm~\cite{Cacciari:2008gp, Cacciari:2011ma} with a distance parameter of 0.4 (AK4 jets) or 0.8 (AK8 jets).
Jet momentum is determined as the vector sum of all particle momenta in the jet, and is found from simulation to be, on average, within 5 to 10\% of the true particle-level jet momentum over the whole \pt spectrum and detector acceptance.
Additional $\Pp\Pp$ interactions within the same or nearby bunch crossings (pileup) can contribute additional tracks and calorimetric energy depositions to the jet momentum.
The charged-hadron subtraction~\cite{CMS:2014ata} and pileup-per-particle identification (PUPPI)~\cite{Bertolini:2014bba, CMS:2020ebo} algorithms are used to mitigate the effect of pileup on AK4 and AK8 jets, respectively.
In the PUPPI algorithm, charged particles identified as originating from pileup vertices are removed, while each neutral particle is assigned a probability weight indicating its likelihood of originating from the primary vertex.
The AK8 jets are then clustered using the output PF candidate list, where the four-momenta of each particle are reweighted accordingly.
For both jet types, jet energy corrections derived from simulation ensure that the average measured jet energies are aligned with those of particle-level jets.
To account for residual differences between data and simulation in the jet energy scale, in situ momentum balance measurements are performed in dijet, photon+jet, $\PZ$+jet, and multijet events, with appropriate corrections applied based on these results~\cite{CMS:2016lmd}.

To identify subjets within AK8 jets for the LJP calibration procedure described in Section~\ref{sec:part_perf}, the reconstructed jet constituents are reclustered using the exclusive-\kt algorithm~\cite{Catani:1993hr, Ellis:1993tq}, which uses a target number of subjets as its stopping criterion rather than a distance parameter.
The exclusive-\kt algorithm was found in Ref.~\cite{CMS:2025eyd} to be more effective at identifying subjets than the Cambridge--Aachen~\cite{Dokshitzer:1997in, Wobisch:1998wt} and anti-\kt algorithms widely used in similar analyses.
Following Ref.~\cite{CMS:2025eyd}, the reclustered subjets have their energies corrected using the same factors applied to the respective AK8 jets, and their \kt clustering history is used to measure their LJP.

Two algorithms are used in this paper to calculate the mass of AK8 jets for comparisons with \taggername.
The first is the soft-drop (SD) algorithm~\cite{Larkoski:2014wba}, with parameters $z = 0.1$ and $\beta = 0$, which prunes soft and collinear radiation from the jet.
The algorithm outputs two subjets, whose invariant mass is taken to be the SD mass (\mSD).
The second is a mass regression algorithm using the \particlenet architecture, which leverages PF candidate and SV information in a GNN to predict the jet mass (\mreg), and has demonstrated improved mass resolution of two-pronged jets compared with the SD mass with the aforementioned parameters~\cite{CMS-DP-2021-017}.
The \textsc{DeepJet} recurrent neural network algorithm~\cite{Bols:2020bkb} is used to identify AK4 jets originating from \PQb quarks, as discussed in Section~\ref{sec:part_perf}.

The missing transverse momentum vector \ptvecmiss is computed as the negative vector sum of the transverse momenta of all the PF candidates in an event, and its magnitude is denoted as \ptmiss~\cite{CMS:2019ctu}.
The \ptvecmiss is modified to account for corrections to the energy scale of the reconstructed jets in the event.

\section{Event simulation}
\label{sec:event_sim}

The \taggername algorithm is trained on simulated signal events containing jets from boosted Higgs boson-like decays and background events with top quark-like or QCD jets.
During training, it is essential that a wide range of jet masses and momenta are included, as described in Section~\ref{sec:tagger_training}, to ensure \taggername does not learn to identify specific resonance masses or \pt ranges, thereby decorrelating the tagger output from jet mass and \pt.
This decorrelation is necessary to search for resonances in the jet mass distribution, where it is important that the tagger does not artificially sculpt the background jet mass distribution to mimic the signal, thereby introducing a potential bias in the search.
Several decorrelation approaches have been proposed in the literature, including adversarial training~\cite{Louppe:2016ylz, Shimmin:2017mfk, CMS:2019dqq}, variable transformation~\cite{Dolen:2016kst}, quantile regression~\cite{Moreno:2019neq}, and dedicated loss function penalty terms~\cite{Kasieczka:2020yyl}.
In this work, we instead rely on the generation of a broad range of signal resonance masses, thus implicitly training the tagger to be less sensitive to specific mass values~\cite{CMS-DP-2020-002}.
Higgs boson-like and top quark-like topologies are simulated using, respectively, a spin-2 resonance \PG decaying to \HH and a spin-0 resonance \PS decaying to \ttbar, with the masses of the \PG, \PH, \PS, and \PQt resonances varied as described below.
We exclusively consider AK8 jets in model training and evaluation.
To simplify the data set generation, all events are simulated under the 2017 data-taking conditions.
However, the performance of \taggername trained using these samples has been verified to be consistent when applied to data collected in 2016, 2017, and 2018.
The algorithm is calibrated separately for each data-taking period and is applied to the full Run 2 dataset.

The \PH jet topology is simulated via $\PG\to\HH$, generated with \MGvATNLO 2.6.5~\cite{Alwall:2014hca} at leading order (LO).
The jet \pt and mass ranges are controlled by varying the \PG mass (\mg) from 600 to 6000\GeV in 100\GeV increments, and the Higgs boson mass (\mh) from 15 to 250\GeV in 5\GeV increments for $\mh<50\GeV$ and 10\GeV increments for $\mh>50\GeV$.
The \PG boson width is set to 1\MeV, while the narrow width approximation (NWA) is applied to the Higgs boson.
The \PH decays are simulated with \JHUGEN 7.1.4~\cite{Bolognesi:2012mm} for \hwwq or $\ell\nu\Pq\Pq$ decay processes, and with \PYTHIA 8.205 for \hbb, \hcc, \hssq, \hqq, and \htata decays.
The ratio of the \PW boson mass (\mw) to \mh is kept fixed to the SM value of 0.64~\cite{ParticleDataGroup:2024cfk} for each \mh point.
This ensures that one of the intermediate \PW bosons is on-shell while the other is off-shell, which improves the mass decorrelation and performance on the SM \hwws topology, as described in Section~\ref{sec:tagger_training}.
While \mw in principle affects the relative branching fractions of the \hwwq and $\ell\nu\Pq\Pq$ decay modes, the two processes are generated separately in equal proportion.

The \PQt jet topology is simulated using $\PS\to\ttbar$, generated with \MGvATNLO 2.6.5 at LO.
The masses of the \PS boson (\ms) and the top quark (\mt) are varied as in the \PH jet setup, with the widths set to 1\% of \ms for the \PS boson and the NWA for the top quark.
The \PW boson decays are simulated at the matrix element level, with the \mw again configured such that $\ifrac{\mw}{\mt} = 0.47$, \ie, at the SM value~\cite{ParticleDataGroup:2024cfk}.
This setup improves the decorrelation of \PQt tagging performance from the jet mass, while maintaining sensitivity to the top quark jet background characteristic of the SM configuration.
Unlike the \hwwq case, where the tagger is optimized specifically for the SM \hwws decay topology, we additionally smear $\ifrac{\mw}{\mt}$.
This is achieved by setting the width of the \PW boson equal to \mw, \ie, significantly larger than its SM value, resulting in the \mw being more varied across events.
This further improves the sensitivity to BSM top quark-like decays where the ratio $\ifrac{\mw}{\mt}$ may be different from the SM value.
The \PG and \PS particles are used purely as convenient means to produce and control the kinematic properties of \PH and \PQt jets and do not otherwise influence \taggername training.
In this section, \PH denotes a Higgs boson-like scalar that decays through the same modes as the SM Higgs boson but whose mass is varied over the range given in Table~\ref{tab:tagger_samples}, and \PQt likewise denotes a top quark-like particle of varying mass.
The symbol \PY is instead reserved for the BSM scalar resonances considered in Section~\ref{sec:tagger_performance}, following the convention of CMS searches for $\PX\to\PH\PY$.
The masses of the particles considered in \PH and \PQt jet simulations are summarized in Table~\ref{tab:tagger_samples}.

The QCD multijet events are simulated at LO using \PYTHIA 8.205~\cite{Sjostrand:2014zea} with the jet \pt sampled from different ranges to ensure broad coverage in jet mass and \pt.
The ``ghost association'' method~\cite{Cacciari:2007fd} is employed to determine the flavor of the partons that produce the jets, wherein reconstructed jets are reclustered with the final generated \PQb or \PQc hadrons in their decay chains.
Their four-momenta are adjusted to preserve only directional information without affecting the jet momentum.
The jet origin is determined by counting the number of ghost-associated \PQb or \PQc hadrons.
Jets originating from resonant boosted Higgs bosons and top quarks are categorized according to the number and flavor of daughter quarks falling within the reconstructed AK8 jet.

Additional independent test samples are generated with SM \hwws and \ttbar jet topologies to evaluate the performance of \taggername for SM searches.
The Higgs boson topology for testing is produced through a similar process to that for the training sample, but with \mg fixed at 1000\GeV and SM values for \mh and \mw.
The SM top quark-like jets are simulated using the same resonant \ttbar process, with \ms ranging from 1200 to 4500\GeV.
The parameters $\mw$, $\mt$, and the \PW boson width are all fixed to their SM values.
Finally, the performance is evaluated for BSM signals of scalar bosons (\PY) decaying to SM \PW bosons as well, for searches for new Higgs boson-like resonances away from the SM \mh.
These are simulated similarly to the \PH jet training samples, with \PG decaying to \HY and \mg and the \PY mass (\my) varying, but the \PY decaying to the SM \PW bosons.

\begin{table}[htb!]
    \renewcommand{\arraystretch}{1.2}
    \centering
    \caption{Summary of particle masses in the \taggername training samples.}
    \label{tab:tagger_samples}
    \begin{tabular}{lclc}
        \multicolumn{2}{c}{\PH jets ($\PG\to\HH$)} & \multicolumn{2}{c}{\PQt jets ($\PS\to\ttbar$)}                                       \\\hline
        \mg                                        & 600--6000\GeV                                  & \ms & 600--6000\GeV                 \\
        \mh                                        & 15--250\GeV                                    & \mt & 15--250\GeV                   \\
        \mw                                        & Fixed to SM ratio: $0.64 \mh$                  & \mw & Fixed to SM ratio: $0.47 \mt$ \\
        \multicolumn{4}{c}{All other particles: SM values}                                                                                \\
    \end{tabular}
\end{table}

{
    \tolerance=10000
    The calibration of \taggername is demonstrated on simulated \HH events in Section~\ref{sec:hhbbvv}.
    Production via the ggF channel is simulated at next-to-LO (NLO) accuracy in perturbative QCD using \POWHEG 2.0~\cite{POWHEG1, POWHEG2, POWHEG3, Bagnaschi:2011tu}, whereas the VBF channel is simulated at LO using \MGvATNLO 2.6.5~\cite{Alwall:2014hca} because the corresponding NLO corrections are small~\cite{Dreyer:2018qbw}.
    A basis of samples is simulated with different \kapvv, \kapl, and \kapv coupling modifiers and linear combinations of these are used to model all possible variations.
    The parameter \kapl is the coupling modifier of the Higgs boson trilinear self-interaction, \kapvv the interaction between two vector bosons and two Higgs bosons, and \kapv the interaction between a Higgs boson and two vector bosons.
    In this work, these additional samples are used to demonstrate the stability of the LJP calibration procedure.
}

The production cross sections of the ggF samples are normalized to the next-to-NLO (NNLO) predictions for each coupling value~\cite{Grazzini:2018bsd, Dawson:1998py, deFlorian:2013jea, deFlorian:2015moa, Baglio:2018lrj, Borowka:2016ehy, Shao:2013bz}, whereas the SM VBF production cross section is normalized to the prediction at next-to-NNLO (N$^3$LO) in QCD~\cite{Dreyer:2018qbw} with the same N$^3$LO/LO correction applied to all other couplings.

For all processes in this paper, the parton showering, hadronization, and underlying event interactions are simulated using \PYTHIA 8.205~\cite{Sjostrand:2014zea} with the CP5~\cite{CMS:2019csb} tune.
The parton distribution functions used correspond to those from NNPDF 3.0~\cite{Ball:2014uwa} and 3.1~\cite{Ball:2017}.
The response of the CMS detector is modeled with the \GEANTfour package~\cite{Agostinelli:2002hh}, with pileup included by overlaying simulated minimum bias interactions on the hard scattering process.

\section{The \texorpdfstring{\taggername}{ParT} algorithm}
\label{sec:tagger}

The \taggername algorithm is an AK8 jet tagger and jet mass regressor that incorporates a dedicated \hwws classification node while achieving decorrelation with the jet mass.
It is a self-attention-based model that is trained to distinguish between QCD jets and a variety of hadronically and leptonically decaying \PH and \PQt jets for general use beyond \hwws identification, as well as to better estimate the jet mass.
This section describes the \taggername model, including training samples, input features, model architecture, training procedures, and performance benchmarks.

\subsection{Model architecture and inputs}
\label{sec:tagger_arch}
We employ a particle transformer architecture, which has been shown to achieve state-of-the-art performance on large-scale data sets~\cite{Qu:2022mxj}, outperforming convolutional and graph neural networks, such as \particlenet, with similar computational efficiencies.
The \partmodel algorithm processes input particles and vertices as a permutation-invariant set of tokens, updating each particle's features through a series of modified multi-head self-attention blocks, referred to as ``particle attention blocks'' (PABs).
Beyond the standard self-attention mechanism~\cite{Vaswani:2017attention}, which allows the model to weigh the importance of different inputs when updating their collective representations, PABs incorporate pairwise features between the input particles and vertices that are embedded and used as attention biases in each block.
After eight PABs, two more class attention blocks (CABs)~\cite{Touvron:2021going} use the learned features to update a randomly initialized class token, which aggregates these features into a global representation of the jet.
This class token is then passed through a final multi-layer perceptron (MLP) that outputs the class probabilities, which are normalized by a softmax function~\cite{NIPS1989_0336dcba}, and the jet mass as an auxiliary training objective (as discussed in Section~\ref{sec:tagger_training}).
In this way, the model employs a unified backbone that jointly performs jet tagging and jet mass regression while maintaining the performance of both tasks.
A diagram of the architecture is shown in Fig.~\ref{fig:part_arch}.

Transformers are also more computationally efficient than GNNs.
The \taggername model has a significantly larger capacity than the \particlenet model of Refs.~\cite{CMS-DP-2020-002, CMS-DP-2021-017}, with 2.3 million parameters compared with 500,000.
However, its inference time is 6\% smaller than that of \particlenet.
The \taggername model has also been shown to require fewer floating point operations per inference than comparable \particlenet models~\cite{Qu:2022mxj}.

\begin{figure}[ht!]
    \centering
    \includegraphics[width=\textwidth]{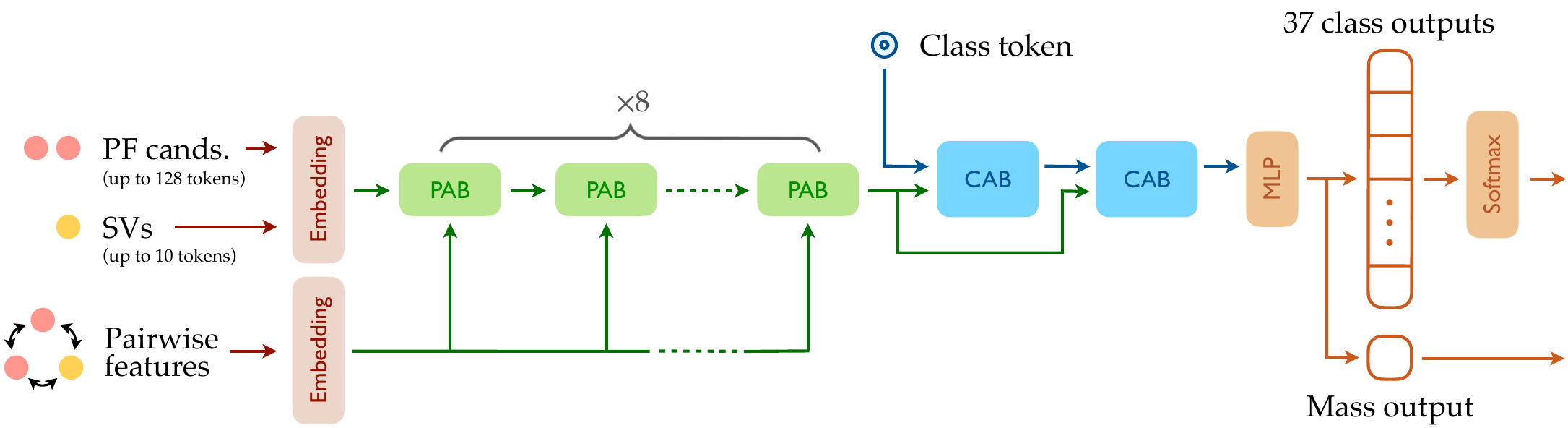}
    \caption{Diagram of the \taggername model architecture.
        The model processes two different sets of input features per jet, from PF candidates and SVs.
        These features are embedded using separate MLPs into 128-dimensional representations before being concatenated and passed through eight PABs.
        Pairwise features are also calculated between each input element, which are embedded using a single one-dimensional convolutional layer and used as attention biases for each PAB.
        Two CABs then use the learned features to update a randomly initialized class token, which aggregates these features into a global representation of the jet.
        Their output is then finally passed through an MLP that outputs the class probabilities, which are normalized by a softmax function, as well as the jet mass.
    }
    \label{fig:part_arch}
\end{figure}

A distinguishing feature of our algorithm with respect to Ref.~\cite{Qu:2022mxj} is the use of two different sets of input features: PF candidates and SVs within the jet.
All jets are used, but the number of input elements per jet is capped.
Based on the distributions of their multiplicities, we select up to 128 PF candidates and 10 SVs per jet.
These limits are chosen to be large enough that, for more than 99.5\% of jets, every reconstructed PF candidate and SV is passed to the network.
For the small fraction of jets whose multiplicities exceed these limits, only the highest-\pt objects are retained.
The PF candidates, including both charged and neutral particles, are taken as constituents of the AK8 jet.
Their input features include kinematic information, particle charge and type, as well as trajectory-related properties in the case of charged particles.
Properties of SVs whose momentum direction lies within the jet, such as kinematic information and impact parameters, are also included as input features.
The full set of features is listed in Table~\ref{tab:part_inputs}.
The features of both data groups are embedded using separate MLPs into 128-dimensional representations before being concatenated and passed through the PABs.
All input features are preprocessed such that their means and standard deviations across the training distributions are approximately 0 and 1, respectively.

Pairwise features $\ln \Delta, \ln (\kt/\GeVns), \ln z,$ and $\ln (m^2/\GeVns^2)$ are calculated between all input elements from their four-momenta~\cite{Dreyer:2020brq}, where
\begin{align}
    \Delta       & = \sqrt{{(\Delta y)^2 + (\Delta \phi)^2}},                                  \\
    k_{\text{T}} & = \min(p_{\text{T},a}, p_{\text{T},b}) \Delta,                              \\
    z            & = \min(p_{\text{T},a}, p_{\text{T},b}) / (p_{\text{T},a} + p_{\text{T},b}), \\
    m^2          & = (E_a+E_b)^2 - \|\mathbf{p}_{a}+\mathbf{p}_{b}\|^2,
    \label{eq:interaction}
\end{align}
$\Delta y$ and $\Delta\phi$ are the pairwise differences in rapidity and the azimuthal angle, respectively;
$p_{\text{T},i} = (p_{x, i}^2+p_{y, i}^2)^{1/2}$ is the transverse momentum;
and $\mathbf{p}_i=(p_{x,i}, p_{y,i}, p_{z,i})$ is the momentum 3-vector and $\|\cdot\|$ is the norm, for $i=a$, $b$.
These features are embedded into a 64-dimensional representation using one-dimensional convolutional layers and then used as attention biases for each PAB.
These architectural hyperparameters are based on those in Ref.~\cite{Qu:2022mxj}.

\begin{table}[htbp!]
    \begin{center}
        \caption{The complete set of input features per AK8 jet used for the \taggername model training. Two types of inputs are considered: PF candidates and secondary vertices (SVs). The PF candidate features marked with a star $(\star)$ apply only to charged PF candidates and a null value is used for neutral candidates.}
        \label{tab:part_inputs}
        \resizebox*{\textwidth}{!}{
            \begin{tabular}{ll}
                Variable                           & Definition                                                                                           \\\hline
                \multicolumn{2}{c}{{PF candidates}}                                                                                                       \\[2mm]
                $\ln(\pt/\GeVns)$                  & Logarithm of the particle \pt                                                                        \\
                $\ln(E/\GeVns)$                    & Logarithm of the particle energy                                                                     \\
                $\Delta\eta(\text{jet})$           & Difference in pseudorapidity between the particle
                and the jet axis                                                                                                                          \\
                $\Delta\phi(\text{jet})$           & Difference in azimuthal angle between the particle
                and the jet axis                                                                                                                          \\
                $\abs{\eta}$                       & Absolute value of the particle pseudorapidity                                                        \\[2mm]
                $q$                                & Electric charge of the particle                                                                      \\[2mm]
                \texttt{isMuon}                    & True if the particle is identified as a muon                                                         \\
                \texttt{isElectron}                & True if the particle is identified as an electron                                                    \\
                \texttt{isChargedHadron}           & True if the particle is identified as a charged hadron                                               \\
                \texttt{isPhoton}                  & True if the particle is identified as a photon                                                       \\
                \texttt{isNeutralHadron}           & True if the particle is identified as a neutral hadron                                               \\[2mm]
                \texttt{pvAssociationQuality}      & Quality of the association of the track to the primary vertices $^{(\star)}$                         \\
                \texttt{lostInnerHits}             & Quality of the track related to missing hits on the pixel layers $^{(\star)}$                        \\
                $\chi^2/$ndof                      & The $\chi^2$ value of the trajectory fit normalized to the number of degrees of freedom $^{(\star)}$ \\
                \texttt{qualityMask}               & Quality of the track $^{(\star)}$                                                                    \\
                $d_z$                              & Longitudinal impact parameter of the track $^{(\star)}$                                              \\
                $d_z$/$\sigma_{d_{z}}$             & Significance of the longitudinal impact parameter $^{(\star)}$                                       \\
                $d_{xy}$                           & Transverse impact parameter of the track $^{(\star)}$                                                \\
                $d_{xy}$/$\sigma_{d_{xy}}$         & Significance of the transverse impact parameter $^{(\star)}$                                         \\
                $\eta_{\text{rel}}$                & Pseudorapidity of the track relative to the jet axis $^{(\star)}$                                    \\
                $p_{\text{T,rel}}$ ratio           & Relative track momentum perpendicular to the jet axis $^{(\star)}$                                   \\
                $p_{\text{par,rel}}$ ratio         & Relative track momentum parallel to the jet axis $^{(\star)}$                                        \\
                $d_{\text{3D}}$                    & Signed three-dimensional (3D) impact parameter of the track~\cite{CMS:2017wtu} $^{(\star)}$          \\
                $d_{\text{3D}}/\sigma_{\text{3D}}$ & Signed 3D impact parameter significance of the track $^{(\star)}$                                    \\
                \texttt{trackDistance}             & Distance between the track and jet axis at the point of closest approach  $^{(\star)}$               \\[3mm]
                \multicolumn{2}{c}{{Secondary vertices within the jet}}                                                                                   \\[2mm]
                $\ln(\pt/\GeVns)$                  & Logarithm of the SV \pt, defined as the \pt of the vector sum of the tracks associated with the SV   \\
                $m_{\text{SV}}$                    & Invariant mass of the tracks associated with the SV                                                  \\
                $\Delta\eta(\text{jet})$           & Difference in pseudorapidity between the SV and the
                jet axis                                                                                                                                  \\
                $\Delta\phi(\text{jet})$           & Difference in azimuthal angle between the SV and the
                jet axis                                                                                                                                  \\
                $\abs{\eta}$                       & Absolute value of the SV pseudorapidity                                                              \\[2mm]
                $N_\text{tracks}$                  & Number of tracks associated with the SV                                                              \\
                $\chi^2/\text{ndof}$               & The $\chi^2$ value of the SV fit normalized to the number of degrees
                of freedom                                                                                                                                \\
                $d_{\text{2D}}$                    & Signed two-dimensional (2D) impact parameter
                of the SV                                                                                                                                 \\
                $d_{\text{2D}}/\sigma_{\text{2D}}$ & Signed 2D impact parameter significance of
                the SV                                                                                                                                    \\
                $d_{\text{3D}}$                    & Signed 3D impact parameter of the SV                                                                 \\
                $d_{\text{3D}}/\sigma_{\text{3D}}$ & Signed 3D impact parameter significance of
                the SV
            \end{tabular}
        }
    \end{center}
\end{table}
\subsection{Model training}
\label{sec:tagger_training}

The final states for each topology are categorized by the number of quarks and leptons per jet, and then further separated by flavor.
Notably, jets from fully hadronic \hwws decays are separated into 4- and 3-pronged jets (\qqqq and \qqq), where the 3-pronged jet category corresponds to the case where one of the four \wws daughter quarks falls outside of the jet cone.
Other Higgs boson decay modes, including $\PH\to\Pq\Pq$ and $\PH\to\tau\tau$, are also targeted for applications to other CMS searches, which will be the subject of future work.
The full set of training classes is shown in Fig.~\ref{fig:part_classes}.
\begin{figure}[ht!]
    \centering
    \includegraphics[width=\textwidth]{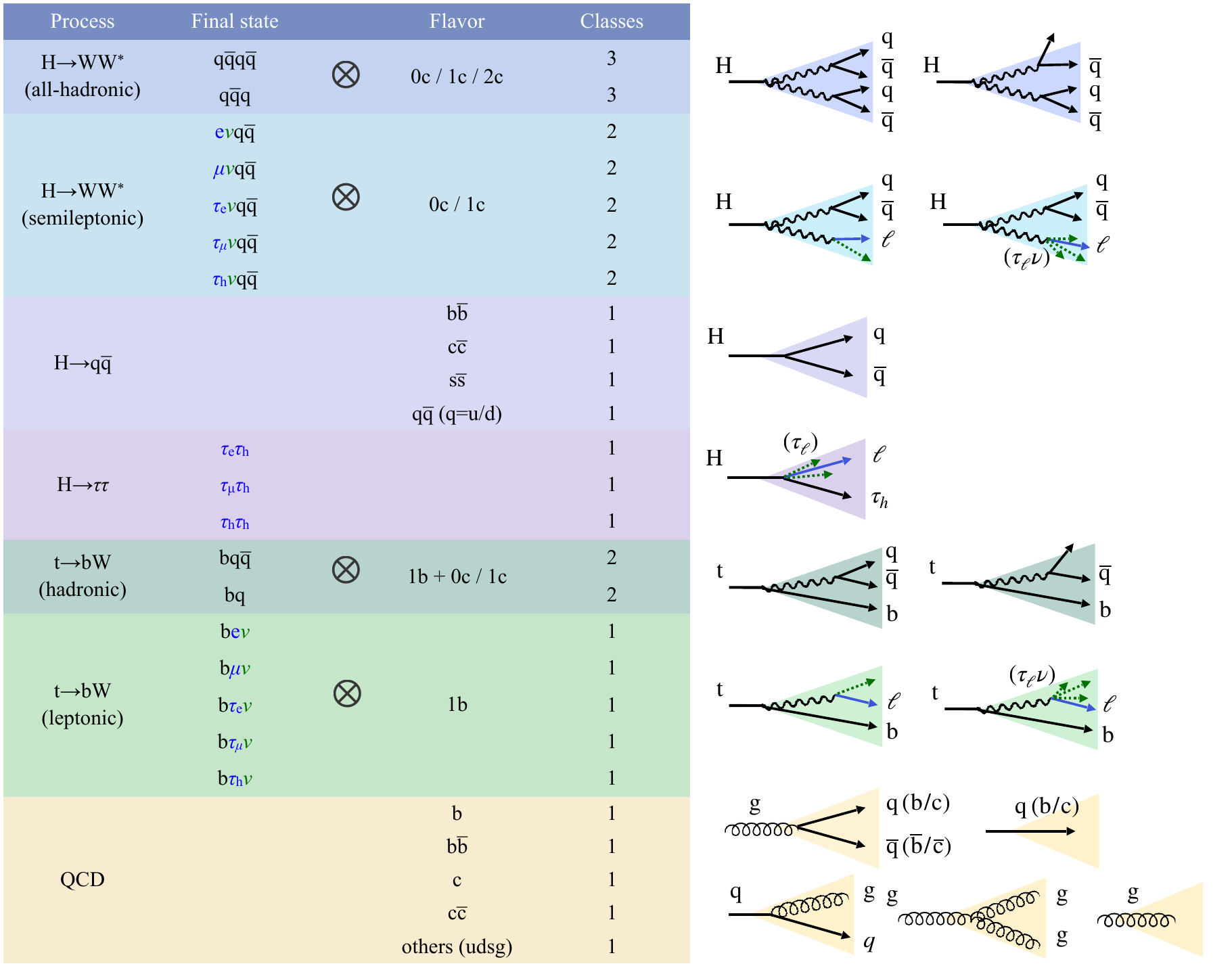}
    \caption{Full suite of AK8 jet topologies considered for the \taggername multiclass classification task.
        Jet types are first categorized by the number of quarks and leptons in the final state, and then further separated by flavor, as shown in the table on the left.
        The symbols $\tau_\Pe$, $\tau_\Pgm$, and $\tauh$ refer to $\tau$ lepton decays to electrons, muons, and hadrons, respectively.
        The total number of subclasses for each process, therefore, is given by the tensor product ($\otimes$) between the different final states and flavors.
        Diagrams illustrating the corresponding jet topologies, which are not exhaustive, are shown on the right.
    }
    \label{fig:part_classes}
\end{figure}

The primary goal of \taggername is jet classification among these 37 classes, along with an auxiliary task of regressing the jet mass.
The classification task uses a cross-entropy loss function, while the mass regression task uses the ``log-cosh'' loss:
\begin{equation}
    \mathcal L_\text{log-cosh}(m_\mathrm{true}, m_\mathrm{pred}) = \ln(\cosh(
    (m_\mathrm{pred} - m_\mathrm{true})/\GeVns)),
\end{equation}
where $m_\mathrm{true}$ is chosen to be the resonance mass, \mh or \mt, for \PH and \PQt jets, and the generator-level \msd for QCD jets, as described in Ref.~\cite{CMS-DP-2021-017}, expressed in units of \GeVns.
The log-cosh function smoothly interpolates between the mean squared error for small differences and mean absolute error for large differences, and has been shown to achieve superior performance for jet mass regression, while being less sensitive to outliers~\cite{CMS-DP-2021-017}.
The overall training objective is to minimize the sum of the cross-entropy and log-cosh losses simultaneously, with the log-cosh loss weighted by a hyperparameter $\lambda$ to balance the two objectives.
The value of $\lambda$ has been tuned empirically to 0.05.

The \taggername algorithm is trained to be decorrelated with the jet mass firstly by varying \mh and \mt in the training samples, as described in Section~\ref{sec:event_sim}.
For \PH decays into light SM particles, \ie, quarks and $\tau$ leptons, varying \mh alone is sufficient to achieve mass decorrelation, while maintaining strong tagging performance, as illustrated by studies of the \particlenet algorithm~\cite{CMS-DP-2020-002, CMS-DP-2021-017}.
However, the high mass of the \PW boson poses a challenge for the \hww topologies.
Namely, if \mw were to remain fixed to its SM value, three distinct kinematic phases would be present in the training samples for the \hww topology as a function of \mh: 1) two off-shell \PW bosons for $\mh \lesssim \mw$; 2) one off-shell and one on-shell \PW boson for $\mw \lesssim \mh \lesssim 2\mw$; and 3) two on-shell \PW bosons for $\mh \gtrsim 2\mw$.
This variation in the jet kinematics was found to harm the model performance on the SM \hwws topology, as well as to introduce biases in the performance with respect to the jet mass.
To address this, training samples are generated with varying \mw as well, as described in Section~\ref{sec:event_sim}, ensuring \taggername is optimized for the SM \hwws on-shell plus off-shell topology and improving its mass decorrelation.

To further improve decorrelation, we ensure training classes share similar mass distributions by sampling uniformly in the two-dimensional (2D) \msd and $\ln(\pt/\GeVns)$ per class, as in Ref.~\cite{CMS-DP-2020-002}.
The 2D sampling weights are derived in 10\GeV bins of \msd between 20 and 260\GeV and 11 bins in \pt uniformly distributed in logarithmic scale between 200 and 2500\GeV, ensuring a uniform distribution in \msd and $\ln(\pt/\GeVns)$.

Additionally, jets from each class are sampled during the training according to the relative weights in Table~\ref{tab:prep-rwgt-classes} to balance the class distributions.
Each of the four major processes: \hwws, $\PH\to\text{2-pronged}$, \PQt, and QCD jets are weighted equally; \ie, the weights in each row in Table~\ref{tab:prep-rwgt-classes} sum to unity.
Within each process, the classes are weighted according to the SM branching fractions, except for $\PH\to\text{2-pronged}$ jets, where the $\bbbar$, $\ccbar$, $\ssbar$, and $\qqbar$ classes are all weighted equally to avoid the \bbbar final state dominating because of its significantly higher branching fraction.
We find that moderate variations in the weighting scheme (\ie, within approximately one order of magnitude) have little impact on the network performance; in particular, networks trained with different weight configurations can be made equivalent by rescaling the output probability scores.
The choice of weights is therefore mainly driven by technical considerations for class balance.

\begin{sidewaystable}[htbp]
    \centering
    \renewcommand\arraystretch{1.2}
    \caption{Relative weights of each of the classes used for training the \taggername model.
        Each of the four major processes: \hwws, $\PH\to\text{2-pronged}$, \tbw, and QCD jets, are weighted equally and have one row dedicated to them each.
    }
    \label{tab:prep-rwgt-classes}
    \resizebox{\textheight}{!}{
        \begin{tabular}{l|ccccccc|ccccccc|ccccccc|c}
            Process     & \multicolumn{7}{c|}{\hwws} & \multicolumn{7}{c|}{$\PH\to$ 2-pronged} & \multicolumn{7}{c|}{\tbw} & QCD                                                                                                                                                                                                                                                                                              \\
            Final state & $\PQq\PQq\PQq\PQq$         & $\PQq\PQq\PQq$                          & $\Pe\PGn\PQq\PQq$         & $\PGm\PGn\PQq\PQq$ & $\tae\PGn\PQq\PQq$ & $\tam\PGn\PQq\PQq$ & $\tauh\PGn\PQq\PQq$ & \bbbar & \ccbar & \ssbar & $\PQq\PAQq$ & $\tae\tauh$ & $\tam\tauh$ & $\tauh\tauh$ & $\PQb\PQq\PAQq$ & $\PQb\PQq$ & $\PQb\Pe\PGn$ & $\PQb\PGm\PGn$ & $\PQb\tae\PGn$ & $\PQb\tam\PGn$ & $\PQb\tauh\PGn$ & All \\\hline
            Weight      & 0.32                       & 0.08                                    & 0.2                       & 0.2                & 0.03               & 0.03               & 0.14                & 0.16   & 0.16   & 0.16   & 0.16        & 0.08        & 0.08        & 0.2          & 0.32            & 0.08       & 0.2           & 0.2            & 0.03           & 0.03           & 0.14            & 1   \\
        \end{tabular}}
\end{sidewaystable}

The training is performed using the \textsc{Weaver} framework~\cite{weaver}, a DL toolkit based on the \textsc{PyTorch}~\cite{PyTorch} Python library, on three Nvidia GTX 3090 GPUs in parallel.
The total data set comprises 52.4 million jets, with 80\% reserved for training and 20\% for validation, and we use a batch size of 512 jets.
As shown in Ref.~\cite{Qu:2022mxj}, training a high-performance transformer-based jet tagging model typically requires tens of millions of jets, preferably approaching 100 million.
The data set size used in this work is sufficient to achieve strong performance while remaining computationally tractable.
Rather than one pass of the entire data set defining an epoch, we define one epoch as processing 15,000 batches or 7.68 million jets (sampled from 80\% of the data set) for training and 1000 batches or 0.512 million jets (sampled from the remaining 20\%) for validation, continuously iterating through the data set.
The model is trained over 50 epochs.
A model snapshot is saved at the end of each epoch, and the model with the lowest hybrid loss on the evaluation set is selected as the final model.
The \textsc{Lookahead} optimizer~\cite{Zhang:2019}, paired with \textsc{RAdam}~\cite{Liu:2019} as the inner optimizer, is used for training.
The initial learning rate is set to $5 \times 10^{-3}$, remaining constant for the first 70\% of iterations before decaying exponentially at the start of each subsequent epoch, and reaching 1\% of the initial value by the final epoch.
Figure~\ref{fig:loss_curve} presents the loss curves during the training for the training and validation data sets, separated into the classification (cross-entropy) and regression (log-cosh) terms.
No evidence of overfitting is observed.

\begin{figure}[htpb]
    \centering
    \includegraphics[width=0.7\textwidth]{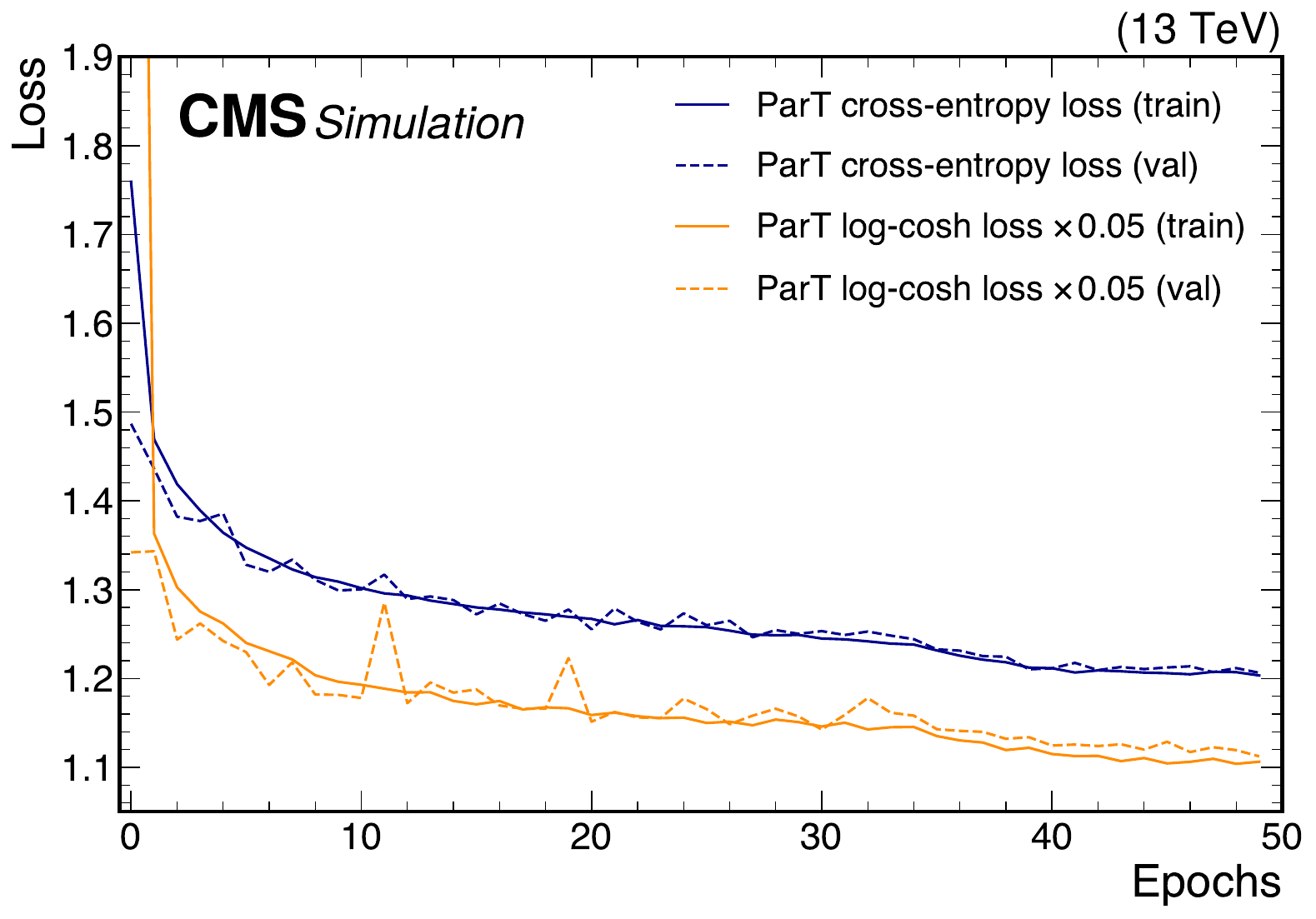}
    \caption{Evolution of the loss function values for the \taggername model on the training and validation data sets over training epochs, shown separately for the classification (cross-entropy) and regression (log-cosh) terms.
    }
    \label{fig:loss_curve}
\end{figure}

\clearpage

\subsection{Performance of jet mass reconstruction}
\label{sec:tagger_mass}

The performance of the \taggername algorithm in reconstructing the jet mass is shown in Fig.~\ref{fig:part_mass} for jets originating from four different processes: \hbb, \hwwq, \tbqq, and QCD, for the SM values of $\mh=125\GeV$, $\mt=172.5\GeV$, and $\mw=80.4\GeV$.
An offline selection of $\pt > 400\GeV$ and $\abs{\eta} < 2.4$ is applied to the AK8 jets, where the \pt selection is more stringent to ensure that the decay products are fully contained within the AK8 jet cone and the full jet mass distribution can be shown up to 250\GeV.
The reconstructed jet mass distributions using the SD and \particlenet algorithms are also shown for comparison.
The \particlenet mass regression algorithm was trained using \hbb, \hcc, \hqq, and QCD jets, with \mh ranging from 15 to 250\GeV~\cite{CMS-DP-2021-017}.
The mass of the resonance peaks $m_{\text{peak}}$ is extracted by locating the maximum of a Gaussian kernel-smoothed mass distribution obtained from each histogram, with the mass resolution quantified by the full width at half maximum (FWHM) of the peak.
Both quantities are shown in the legend of Fig.~\ref{fig:part_mass} for each algorithm and resonant process.

We observe improved mass resolution from both \particlenet and \taggername with respect to \msd for the two-pronged \hbb decay, while \taggername in particular is the most performant for the higher-pronged \hwwq and \tbqq decays.
This is expected because the SD algorithm was not specifically optimized for these 4- and 3-pronged decays, and \particlenet was also not trained on these decays.
One feature of the SD algorithm shown in Fig.~\ref{fig:part_mass}, using the CMS parameters described in Section~\ref{sec:event_reco}, is a relatively strong removal of PF candidates from the jet.
This yields an \msd value close to 0\GeV for a significant fraction of \PH and \PQt jets, and a large tail in the \msd distribution below the true mass.
Despite this, the resolution and scale of \msd have been verified and calibrated with 3-pronged \PQt jets in Run 2 and Run 3 data~\cite{CMS-DP-2023-044,CMS-DP-2025-052}.
We observe a slight bias towards a higher predicted reconstructed mass for \PH and \PQt jets than the true values from both \particlenet and \taggername, which we attribute to the presence of high-mass jets in the training sample.
Finally, we observe smoothly falling distributions for QCD jets for all three algorithms.

\begin{figure}[htpb]
    \centering
    \includegraphics[width=0.49\textwidth]{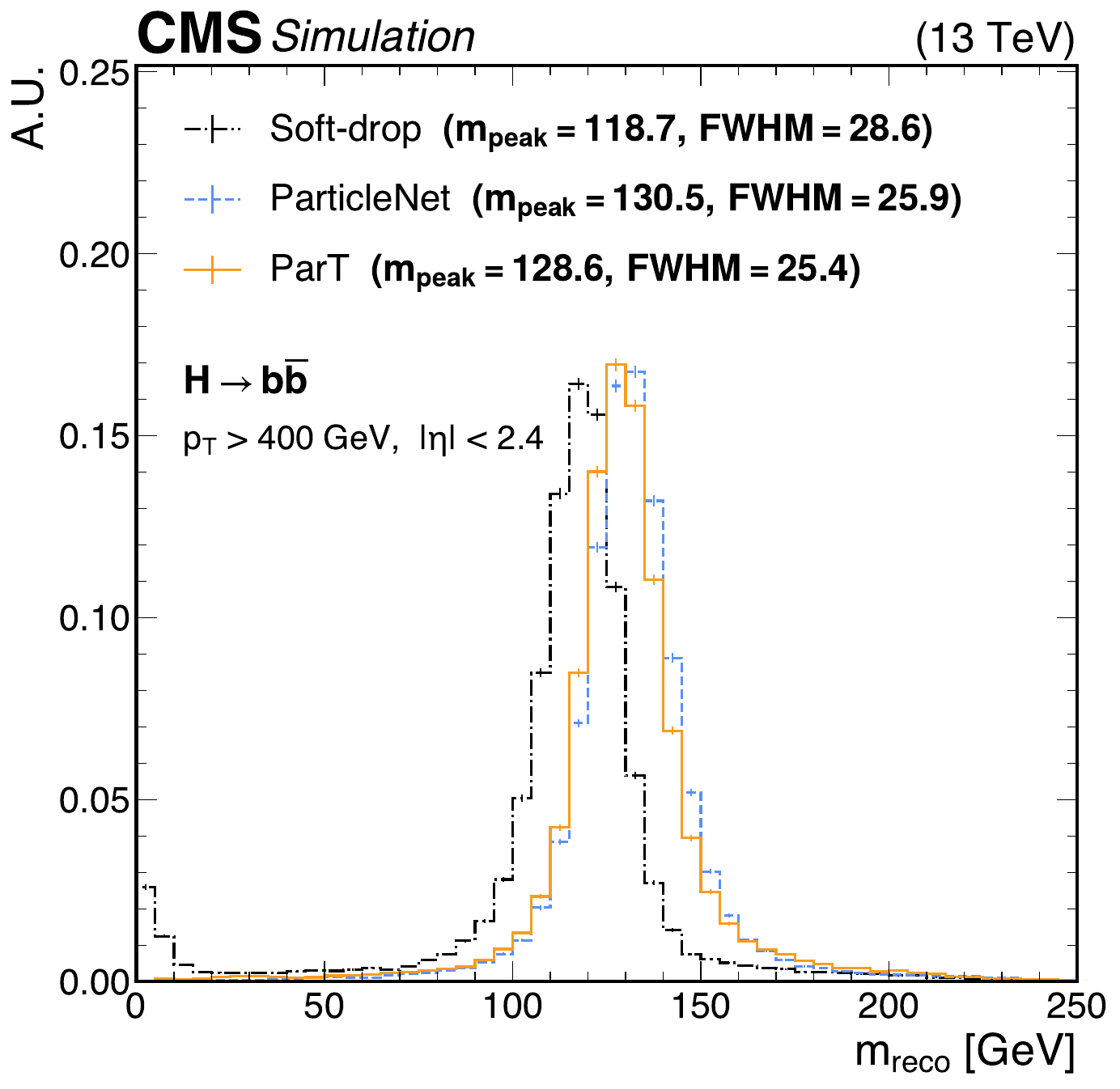}
    \includegraphics[width=0.49\textwidth]{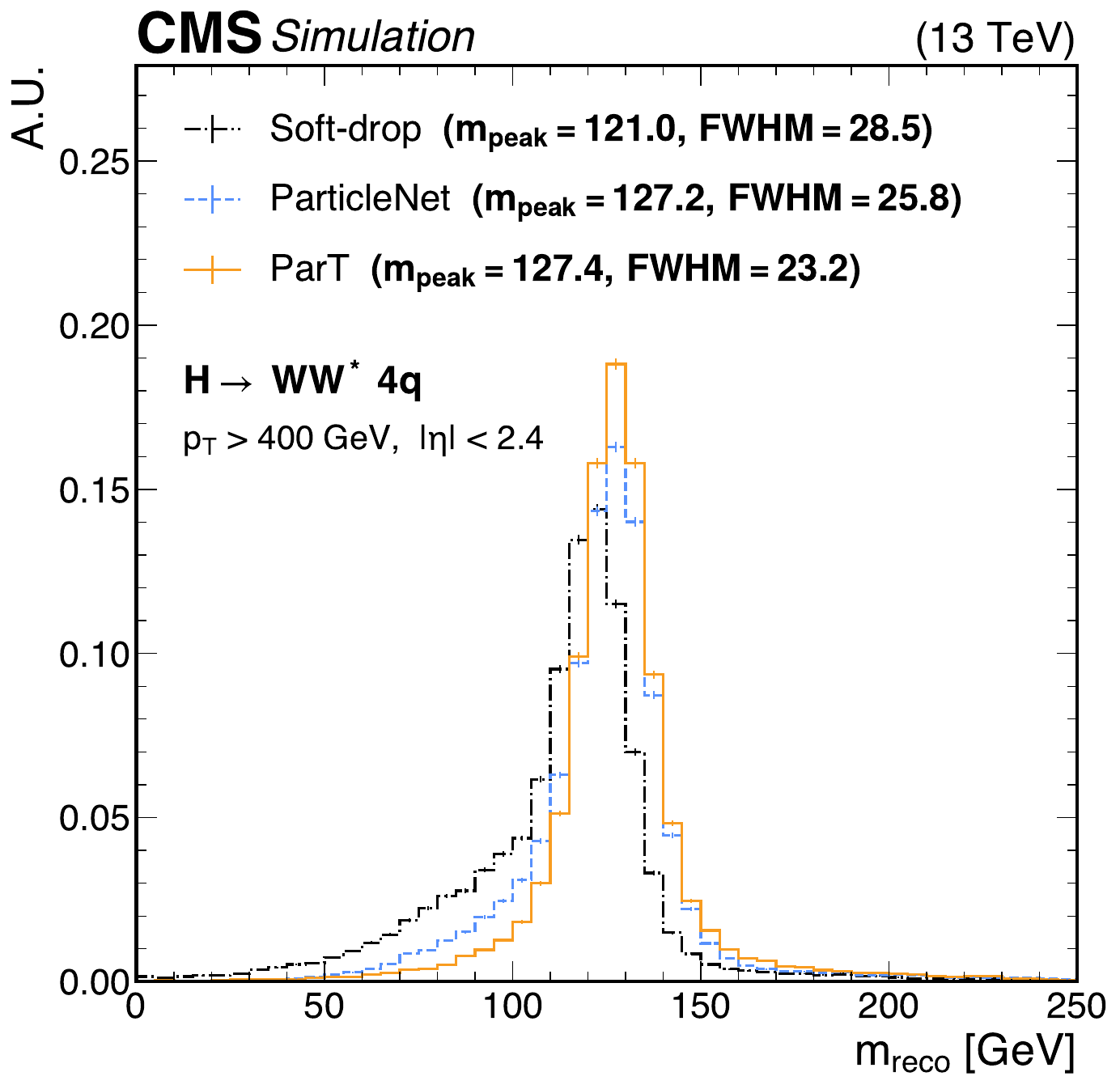}
    \includegraphics[width=0.49\textwidth]{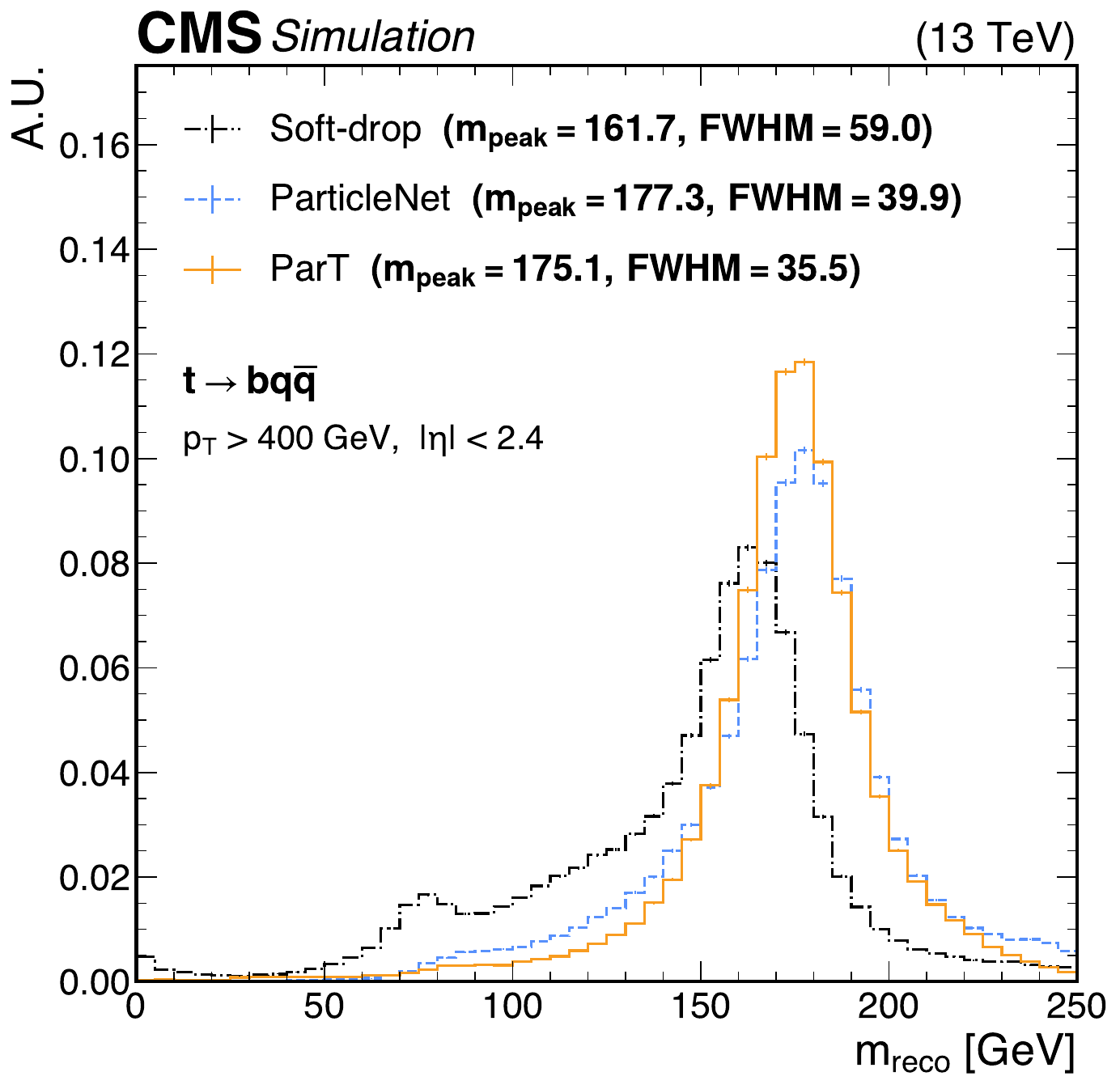}
    \includegraphics[width=0.49\textwidth]{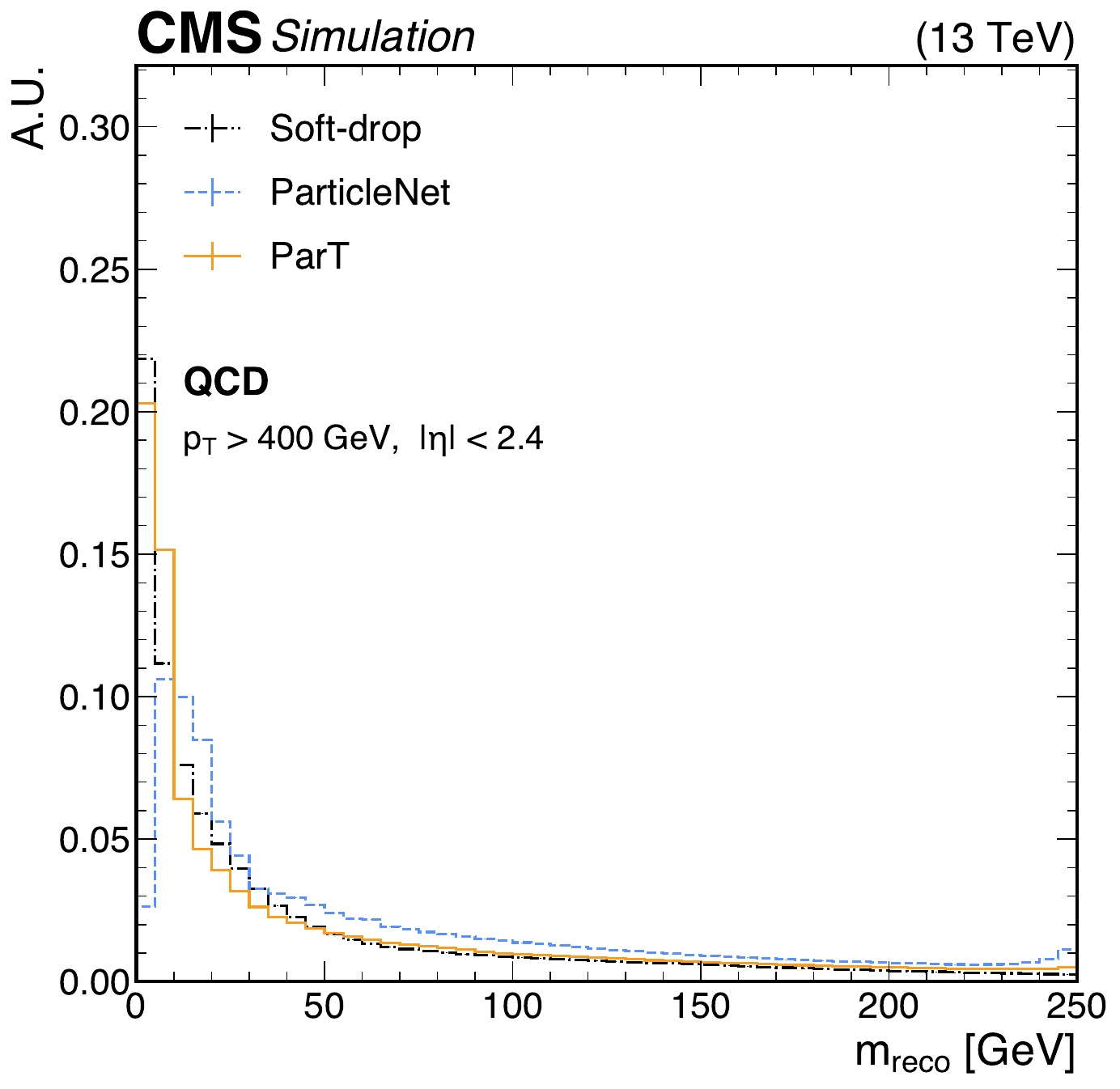}
    \caption{Comparison of jet mass reconstruction (\mreco) using the SD, \particlenet, and \taggername algorithms, for \hbb (upper left), \hwwq (upper right), \tbqq (lower left), and QCD (lower right) jets with the SM values of \mh and \mt.
        An offline selection of $\pt > 400\GeV$ and $\abs{\eta} < 2.4$ is applied to the AK8 jets.
        Statistical uncertainties in the bin yields originating from the limited number of simulated events are represented by vertical error bars.
        The mass at the peak (\mpeak) for each algorithm, calculated using Gaussian kernel density estimation, and the mass resolution, quantified by the FWHM of the resonance peak, are shown as well for \PH and \PQt jets.
    }
    \label{fig:part_mass}
\end{figure}

\section{Performance and calibration of \texorpdfstring{\hwws}{H to WW*} tagging}
\label{sec:part_perf}

\subsection{Performance}
\label{sec:tagger_performance}

To evaluate the performance of the \hwws tagging, we use a single discriminant focusing on distinguishing hadronic \hwws jets from \PQt and QCD multijet jets, defined as
\begin{equation}
    \label{eq:thww}
    \THWW = \frac{\PHWWqqqq + \PHWWqqq}{\PQCD + \PTop + \PHWWqqqq + \PHWWqqq},
\end{equation}
where \PHWWqqqq, \PHWWqqq, \PTop, and \PQCD are the sum of the predicted probabilities of their respective subcategories.

The performance of the tagger using this discriminant in differentiating between SM \hwws ``signal'' and QCD and \PQt background jets is shown in the receiver operating characteristic curves in Fig.~\ref{fig:part_roc}, where signal and background efficiencies are defined as the fraction of signal and background jets passing a given discriminant threshold.
The curves are provided for several ranges of jet \pt.
For the \PH and \PQt jets, we use only the AK8 jets that lie within $\Delta R < 0.8$ of the generator-level \PH or \PQt, respectively, while for QCD we select the jet with the highest \pt in the event.
An offline selection of $\pt > 200\GeV$, $\msd > 30\GeV$, and $\abs{\eta} < 2.4$ is applied to all jets.

We compare our model with the mass-decorrelated \deepak algorithm~\cite{CMS:2020poo}, which was previously state-of-the-art in CMS for \hwws jet identification.
Unlike the \taggername algorithm, which respects inductive biases of jets such as permutation invariance and heterogeneity of the jet constituents, the \deepak algorithm represents jets as lists of PF candidates and SVs and uses a convolutional neural network architecture inspired by \textsc{ResNet}~\cite{resnet} for classification.
The total number of parameters in the \deepak model is about 500,000, and the mass decorrelation is achieved with an adversarial training approach~\cite{Louppe:2016ylz}.
The mass-decorrelated \particlenetmd algorithm was only trained to distinguish between \hbb, \hcc, \hqq, and QCD jets~\cite{CMS-DP-2020-002}, so it cannot be used as a baseline for \hwws jet identification.

The \deepak algorithm was trained on approximately 40 million jets to distinguish among jets originating from $\PH\to4\Pq$ (output \PHqqqq), $\PW\to\qqbar$ (\PWqq), $\PQt\to \PQb\qqbar$ (\PTop) decays, and others.
A discriminant targeting $\PH\to4\Pq$ and $\PW\to\qqbar$ jets, while vetoing QCD jets, can be constructed from its outputs, as in Refs.~\cite{CMS:2021mjl,CMS:2022lqh}.
To further improve the suppression of the \ttbar background, the \deepak discriminant can be extended to veto both QCD and $\PQt\to \PQb\qqbar$ jets.
The resulting discriminant, referred to as the \deepak score, is defined as
\begin{equation}\label{eq:deepak}
    \deepak \text{\ score} = \frac{\PWqq + \PHqqqq}{\PQCD + \PTop + \PWqq + \PHqqqq}.
\end{equation}
The performance of this discriminant is shown in Fig.~\ref{fig:part_roc} as well.
We observe up to ten times stronger QCD jet rejection for the same signal efficiencies using \taggername with respect to the \deepak score, and up to eight times stronger \PQt jet rejection.
Thus, through the architectural and training advancements described in this paper, \taggername significantly outperforms \deepak.

We additionally evaluate the performance of \taggername in identifying BSM scalar resonances \PY decaying to SM \PW bosons, shown in Fig.~\ref{fig:part_yww_roc}.
We observe that, despite being optimized exclusively for the SM $\mw/\mh$ ratio, \taggername performs similarly for a small range of {\my} away from the SM \mh; however, there is some performance degradation for $\my=250\GeV$, at the boundary of the training mass range.
This suggests that \taggername can be used to search for general BSM $\PY \to \PW\PW \to 4\Pq$ signatures for $\my < 250\GeV$~\cite{CMS:2026shw}.
The 15--250\GeV range of resonance masses in the training samples was primarily chosen to decorrelate the tagger response from the jet mass for resonances near the SM Higgs boson mass.
For $\my > 250\GeV$, \taggername is therefore applied outside the mass range covered in training.
The degradation seen at the boundary continues: in the $\PX\to\PH\PY$ search of Ref.~\cite{CMS:2026shw}, the selection efficiency for signal jets with \my above this range is lower by about a factor of three than for those with \my within it.
Extending the training mass range would benefit searches targeting heavier resonances, and is left for future work.
Finally, Fig.~\ref{fig:part_confusion} shows the confusion matrix, with each row indicating the fraction of jets per category classified as the column category by \taggername.
We observe generally strong agreement between the true and predicted classes, with higher misclassification rates within subcategories with similar final states, such as $\hwws\to\tau_\Pe\nu\Pq\Pq$ and $\hwws\to\Pe\nu\Pq\Pq$, as expected.

\begin{figure}[htpb]
    \centering
    \includegraphics[width=0.49\textwidth]{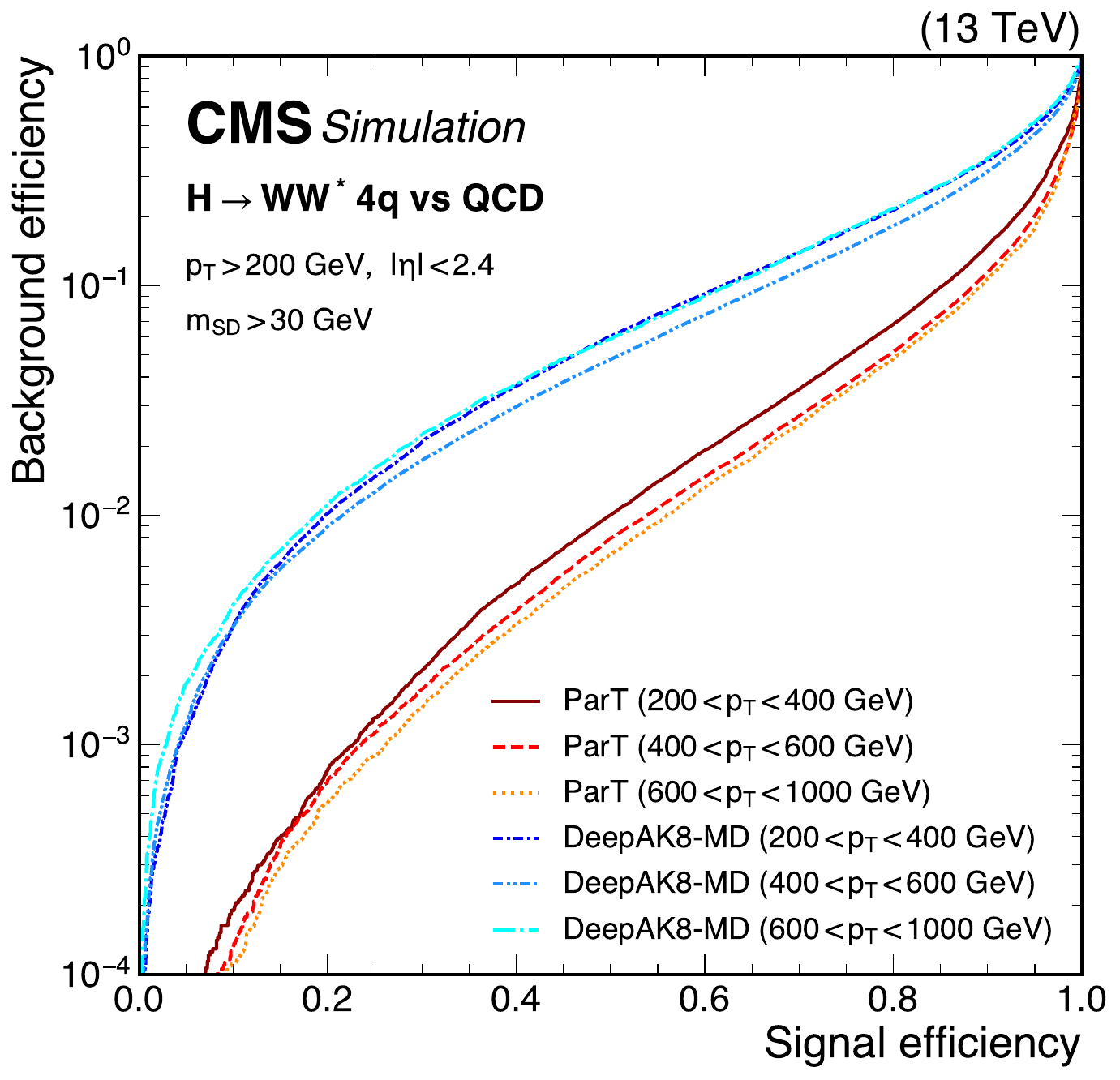}
    \includegraphics[width=0.49\textwidth]{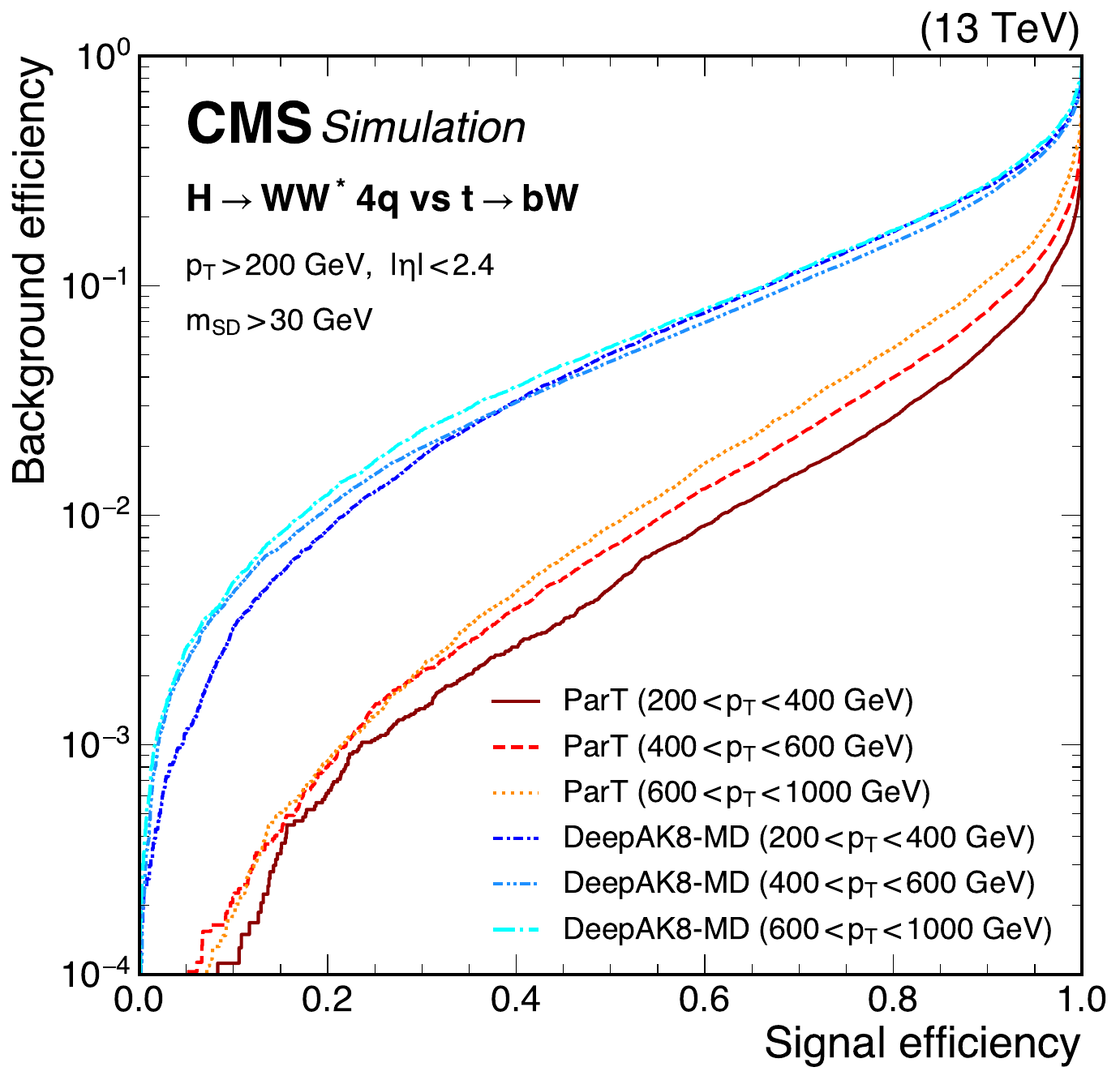}
    \caption{Receiver operating characteristic curves for \hwws signal jets, with the SM values of \mh and \mt, versus background jets from simulated QCD multijet (left) and \ttbar events (right), for the \taggername \THWW and the \deepak scores in the \pt ranges 200--400, 400--600, and 600--1000\GeV.
        An offline selection of $\pt > 200\GeV$ and $\abs{\eta} < 2.4$ is applied to the AK8 jets.
        Signal jets are required to contain all four generator-level quarks from the \PW boson decays within $\Delta R(\text{jet},\,\PQq) < 0.8$.
    }
    \label{fig:part_roc}
\end{figure}

\begin{figure}[htpb]
    \centering
    \includegraphics[width=0.49\textwidth]{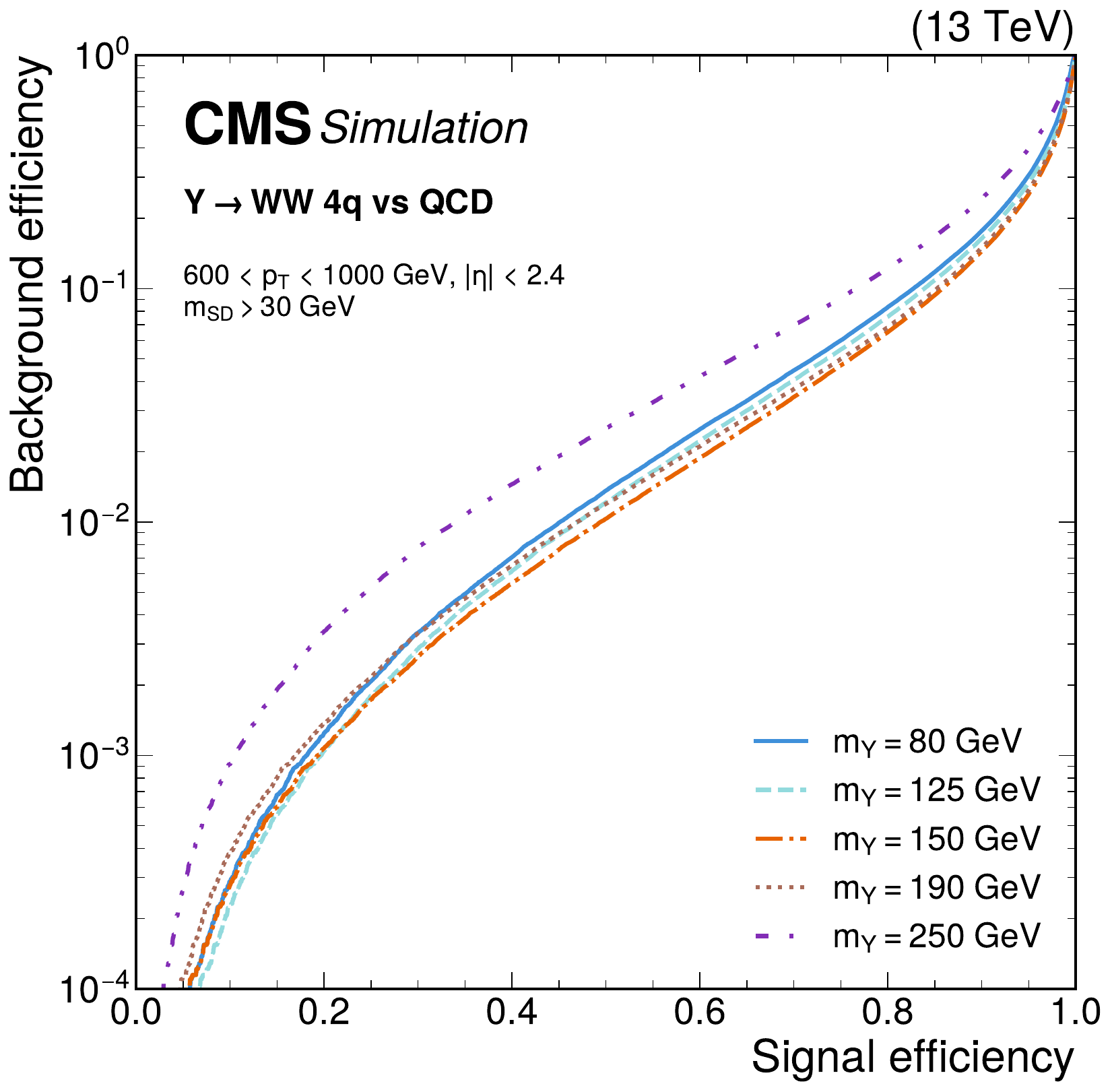}
    \includegraphics[width=0.49\textwidth]{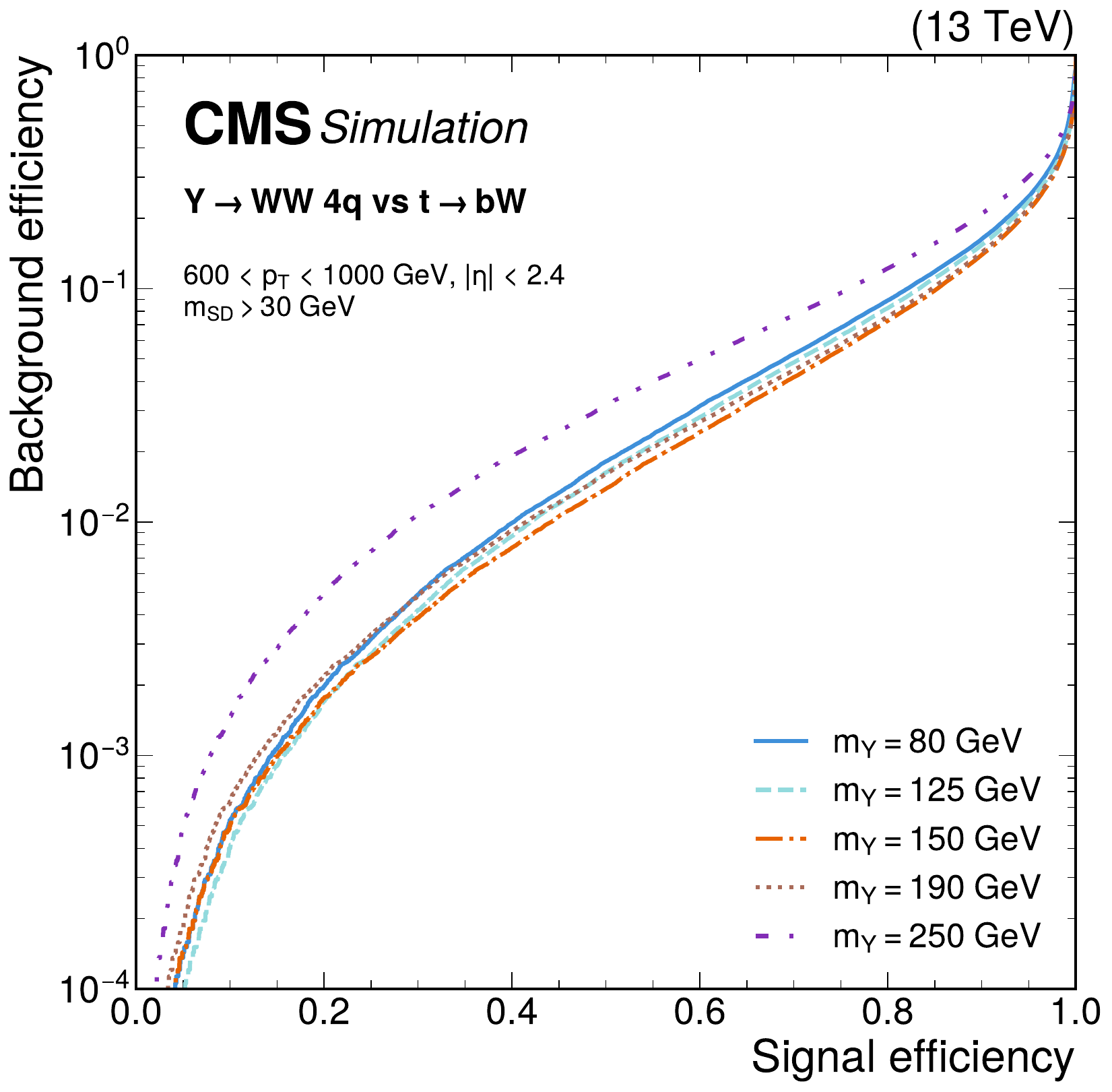}
    \caption{Receiver operating characteristic curves for \yww signal jets, with varying \my and SM \mw, versus background jets from simulated QCD multijet (left) and \ttbar events (right), for the \taggername \THWW score.
        An offline selection of $600 < \pt < 1000\GeV$, $\abs{\eta} < 2.4$, and $\msd > 30\GeV$ is applied to the AK8 jets.
    }
    \label{fig:part_yww_roc}
\end{figure}

\begin{figure}[htpb]
    \centering
    \includegraphics[width=\textwidth]{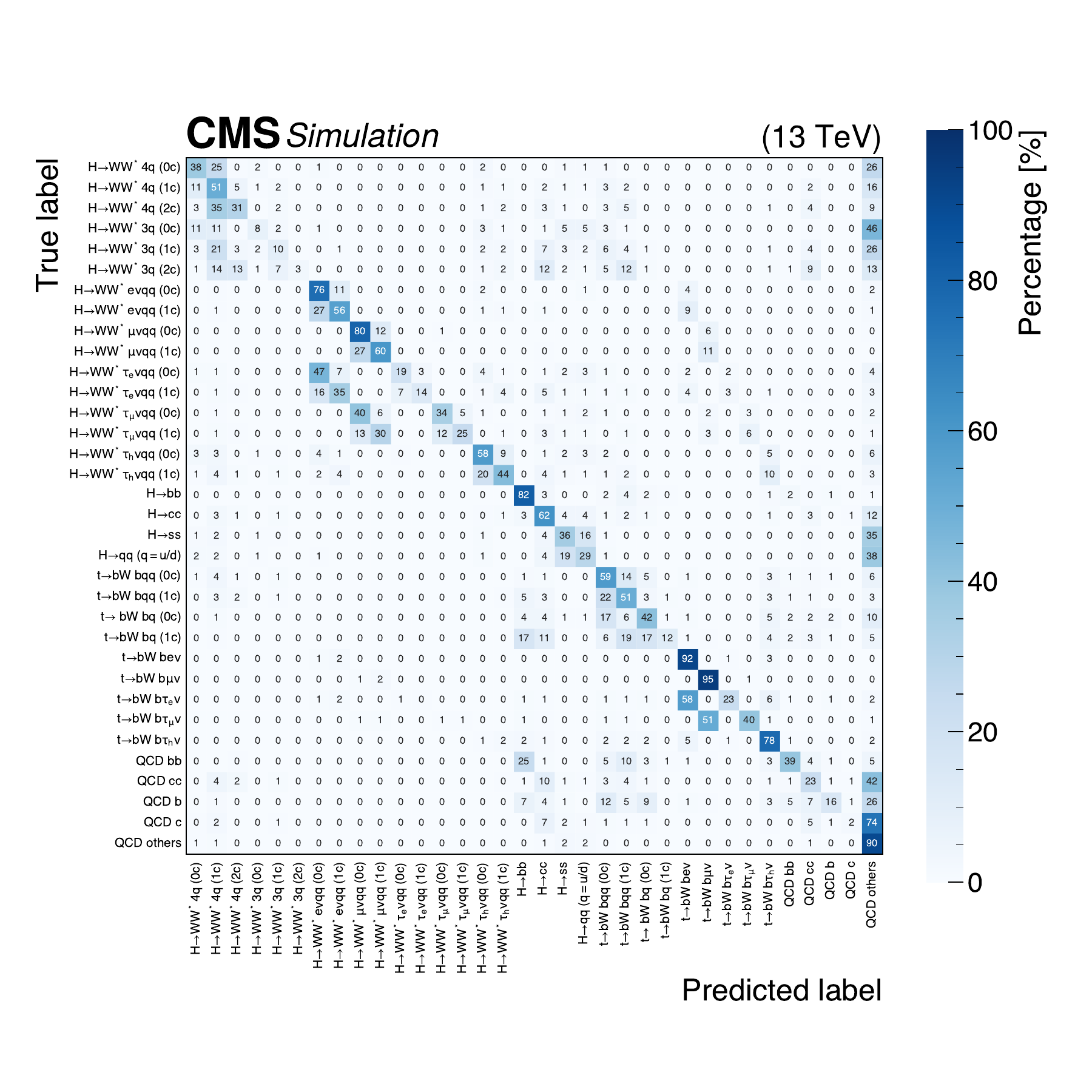}
    \caption{Confusion matrix with each row indicating the fraction of jets per category classified as the column category by \taggername.
        An offline selection of $\pt > 200\GeV$ and $\abs{\eta} < 2.4$ is applied to the AK8 jets.}
    \label{fig:part_confusion}
\end{figure}

\subsection{Mass decorrelation}

In many analyses, it is desirable for the tagger to be decorrelated with the jet mass, as this allows for the use of sideband regions in the jet mass distribution to constrain backgrounds and reduce systematic uncertainties~\cite{Dolen:2016kst}.
The decorrelation of \taggername's outputs with the jet mass is examined in Fig.~\ref{fig:part_md} for jets in simulated QCD multijet events.
Figure~\ref{fig:part_md} also presents the performance of the \deepak discriminant for comparison, which uses an adversarial training strategy for mass decorrelation.
To quantify the degree of mass decorrelation, we consider the Jensen--Shannon distance (JSD)~\cite{jsdref} using base 2, a symmetrized variant of the Kullback--Leibler divergence~\cite{Kullback51klDivergence}, between the \msd distributions with and without \taggername and \deepak selections.
The JSD quantifies the difference between the two distributions and is bounded between 0 and 1, with a lower value indicating better agreement and, hence, better decorrelation.
It is reported in Fig.~\ref{fig:part_md_jsd} for a range of different tagger selection efficiencies and jet \pt bins.

We observe that the \msd distributions remain generally consistent within statistical uncertainties across increasingly tighter selections on the \taggername \THWW discriminant; however, there is some evidence of discrepancies in the tails, motivating an increase of the training mass range.
This will be the subject of future work.
Furthermore, by using samples of varying resonance masses for \taggername rather than the adversarial approach of \deepak, we are able to employ a simpler loss function and obviate the challenge of training a model with competing objectives; thus, we observe consistently lower JSD values for \taggername across a range of selection efficiencies and jet \pt.

\begin{figure}[htpb]
    \centering
    \includegraphics[width=0.40\textwidth]{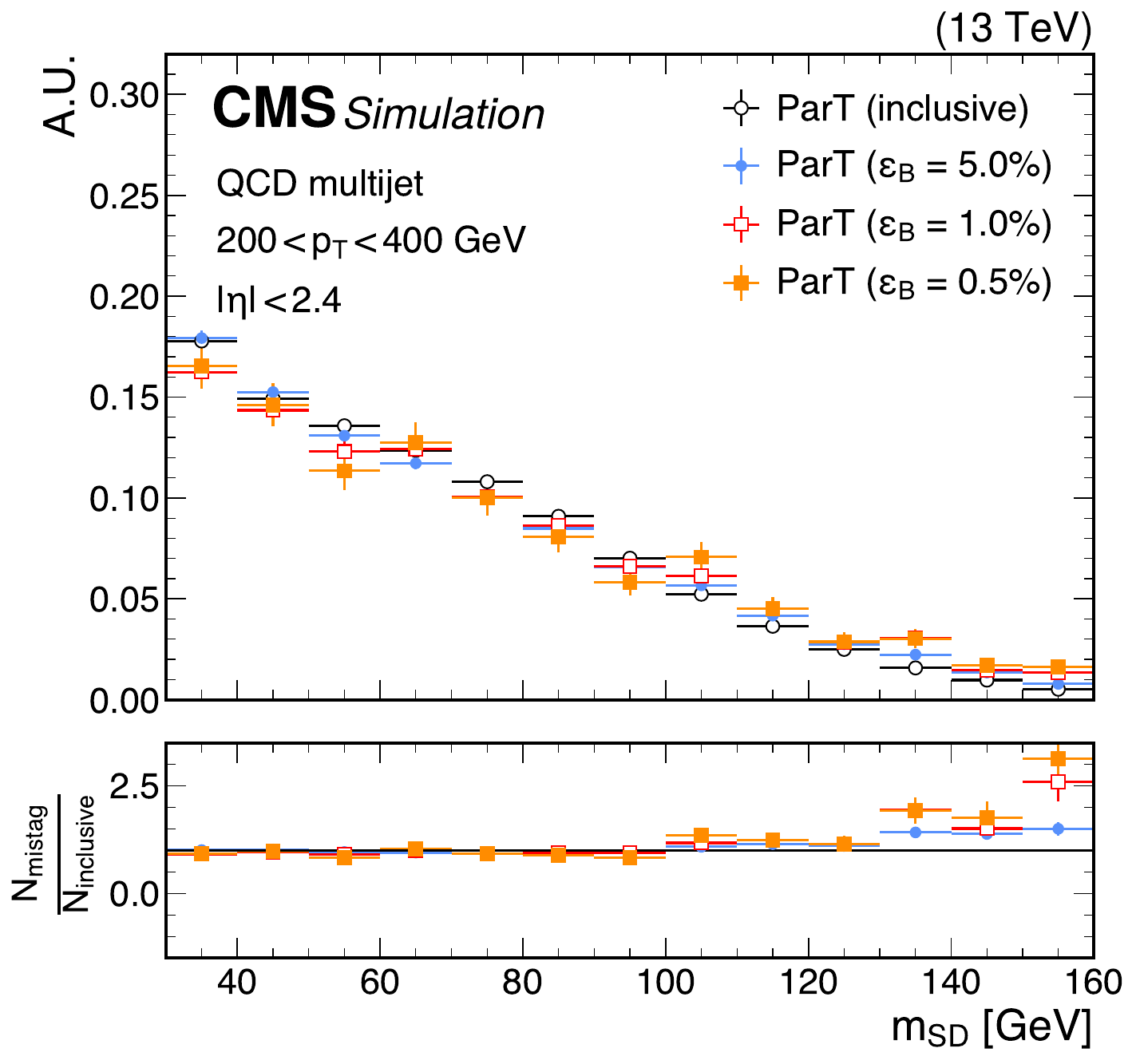}
    \includegraphics[width=0.40\textwidth]{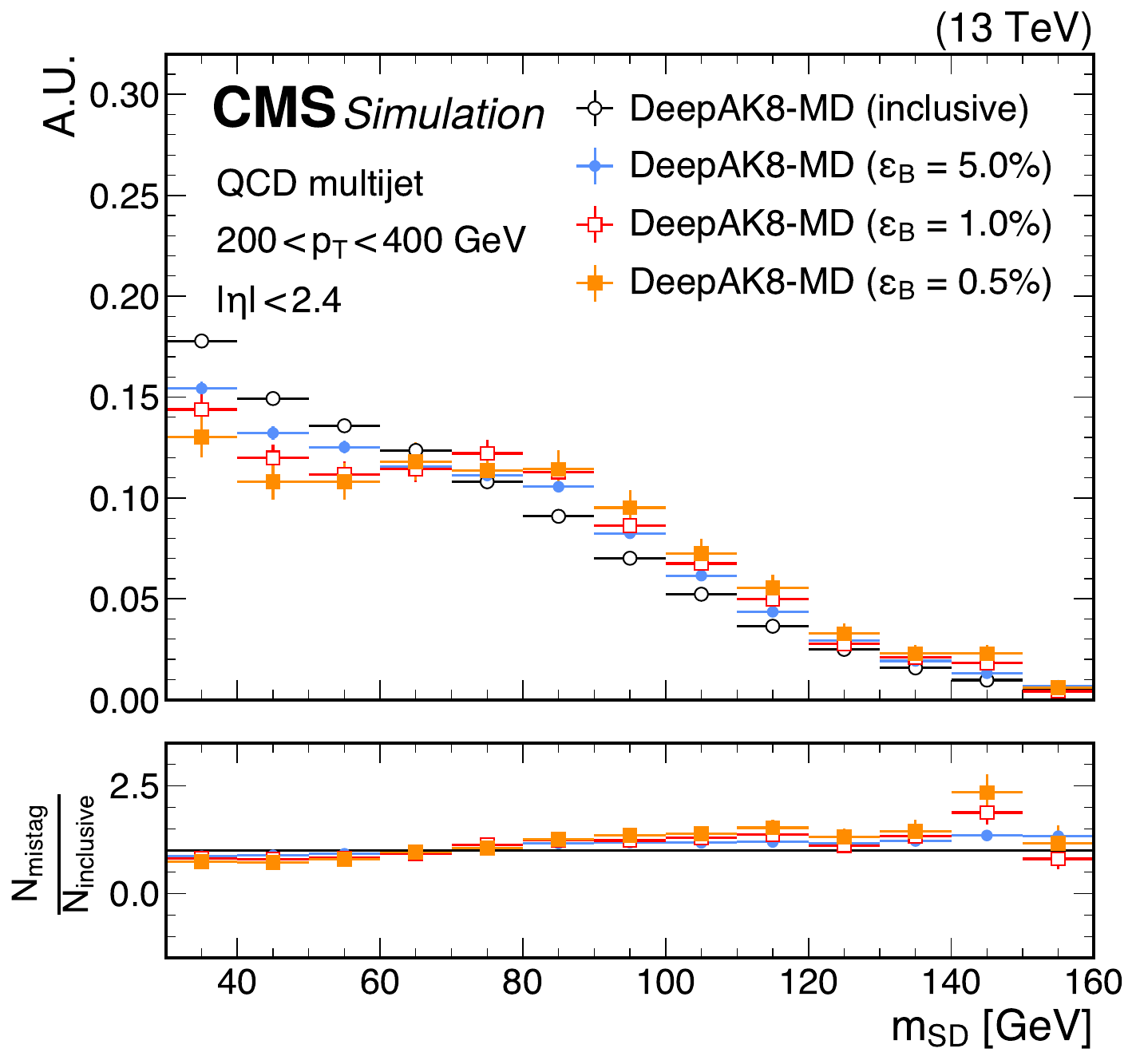}\\
    \includegraphics[width=0.40\textwidth]{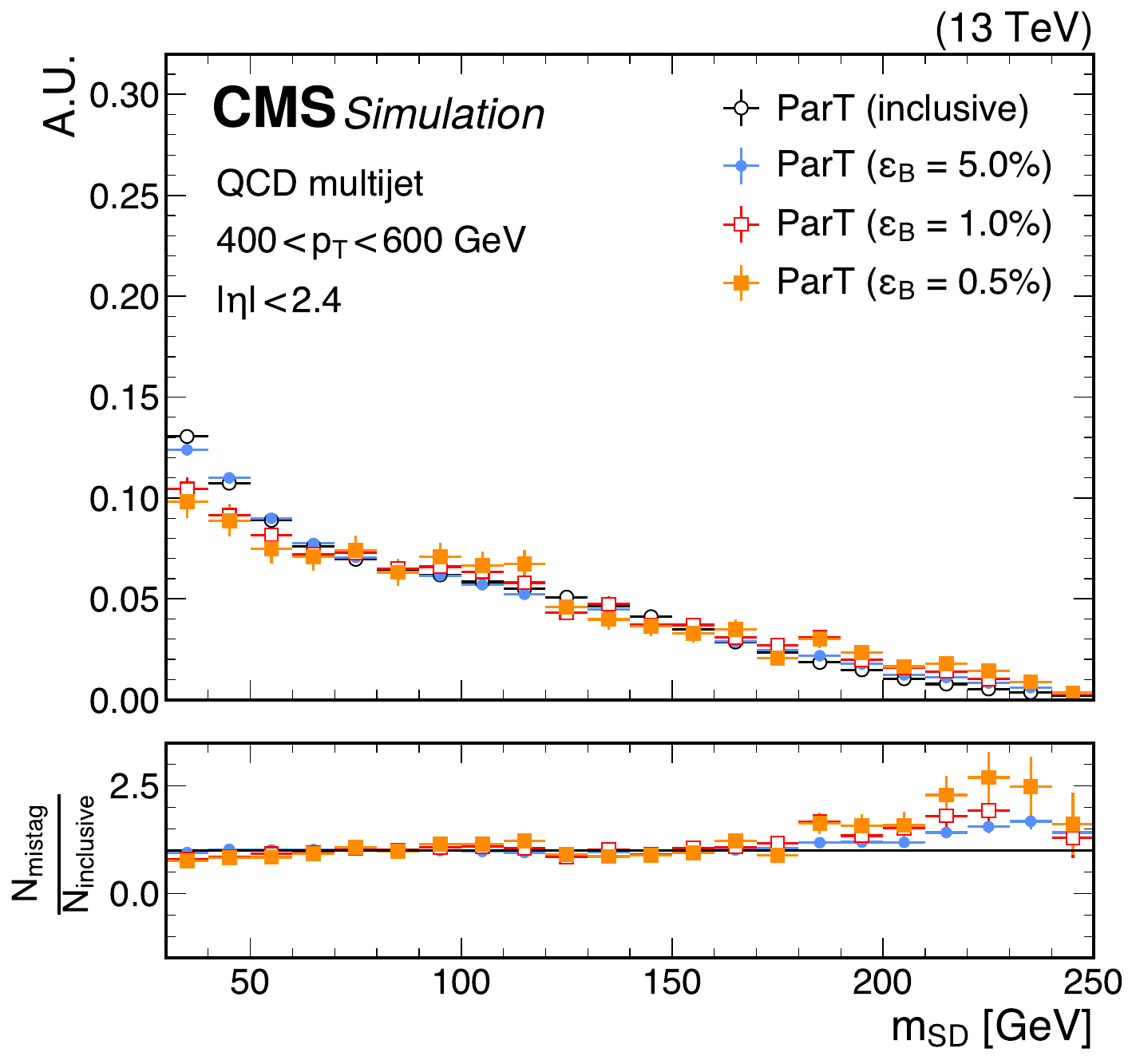}
    \includegraphics[width=0.40\textwidth]{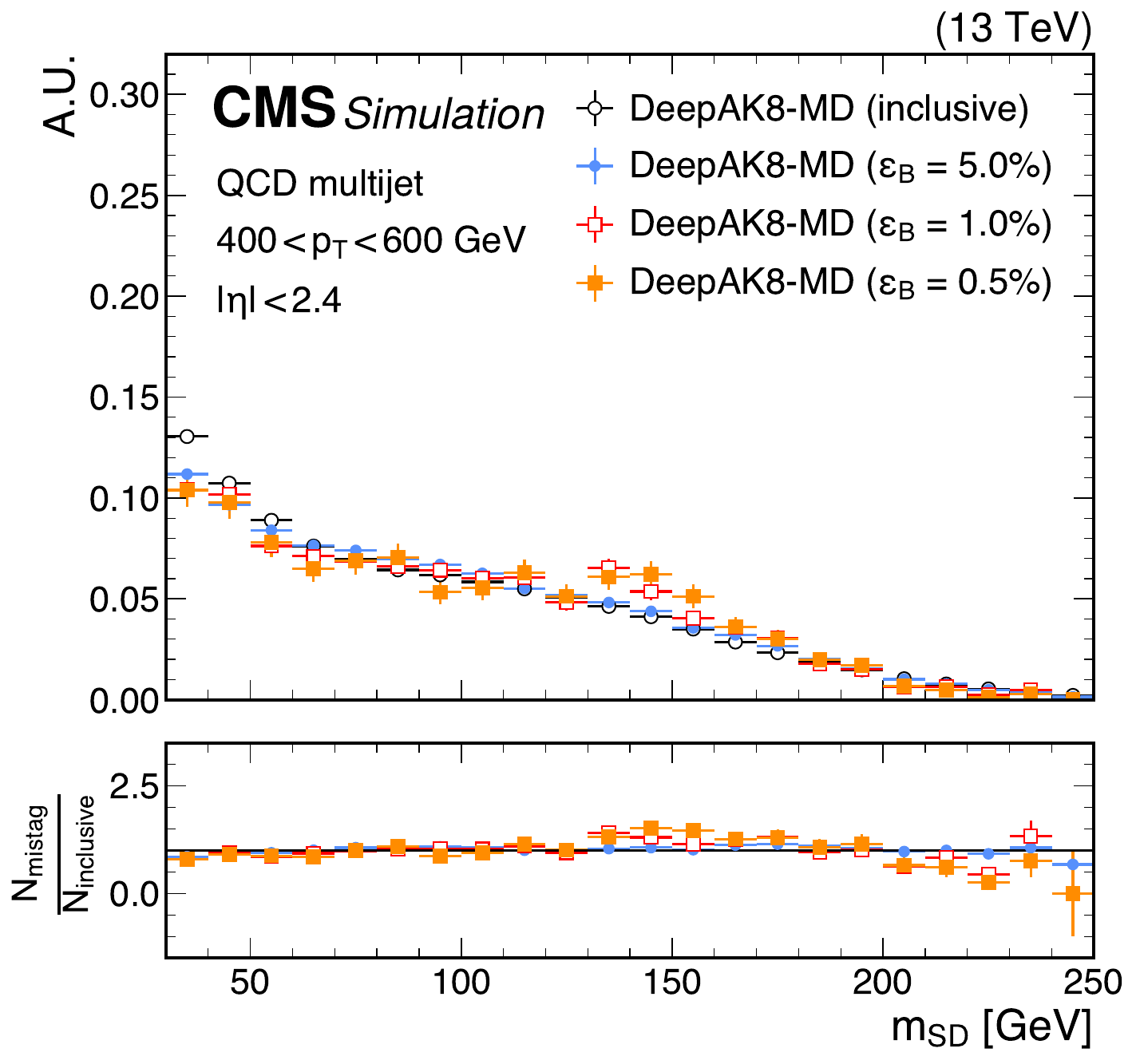}\\
    \includegraphics[width=0.40\textwidth]{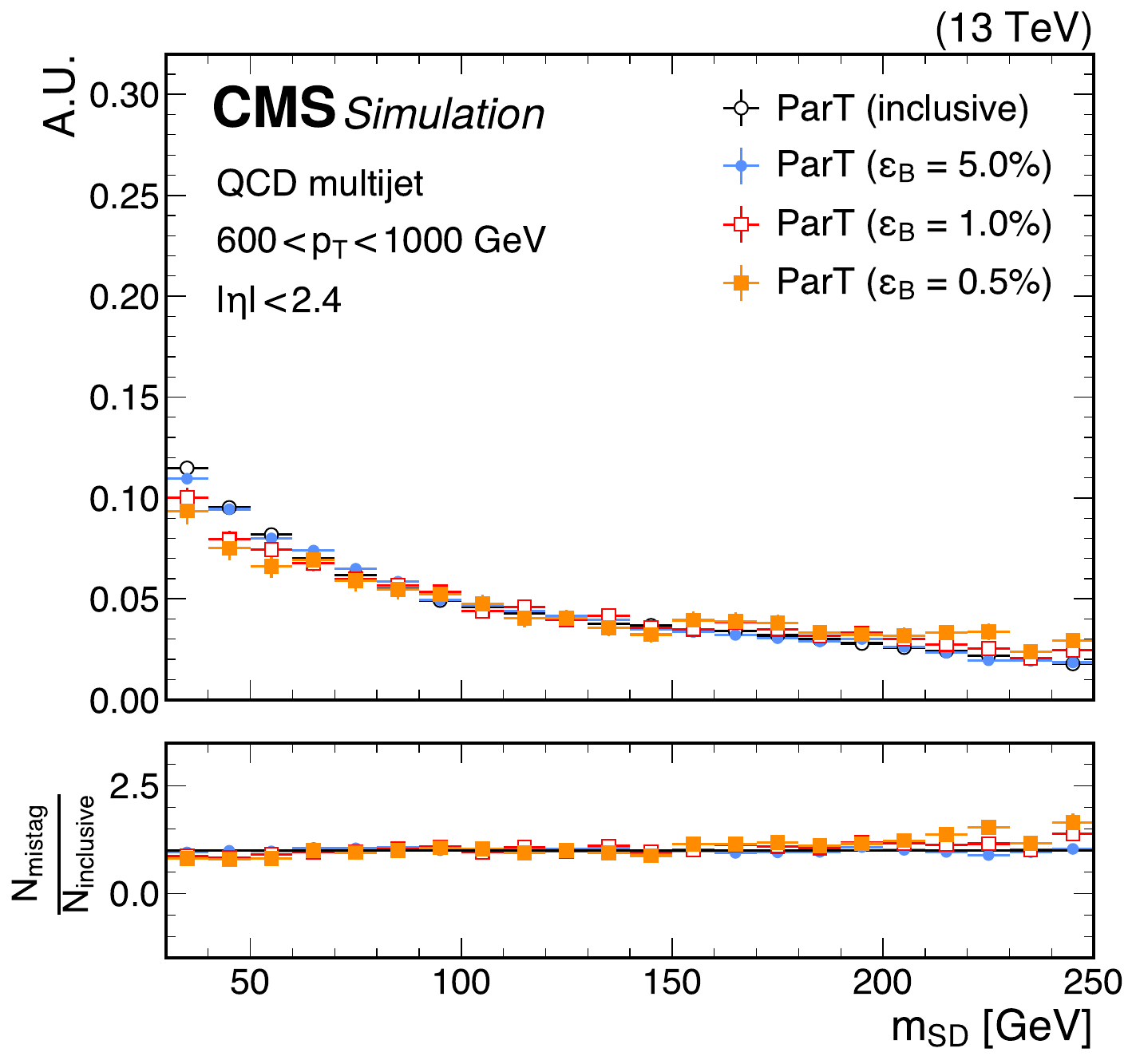}
    \includegraphics[width=0.40\textwidth]{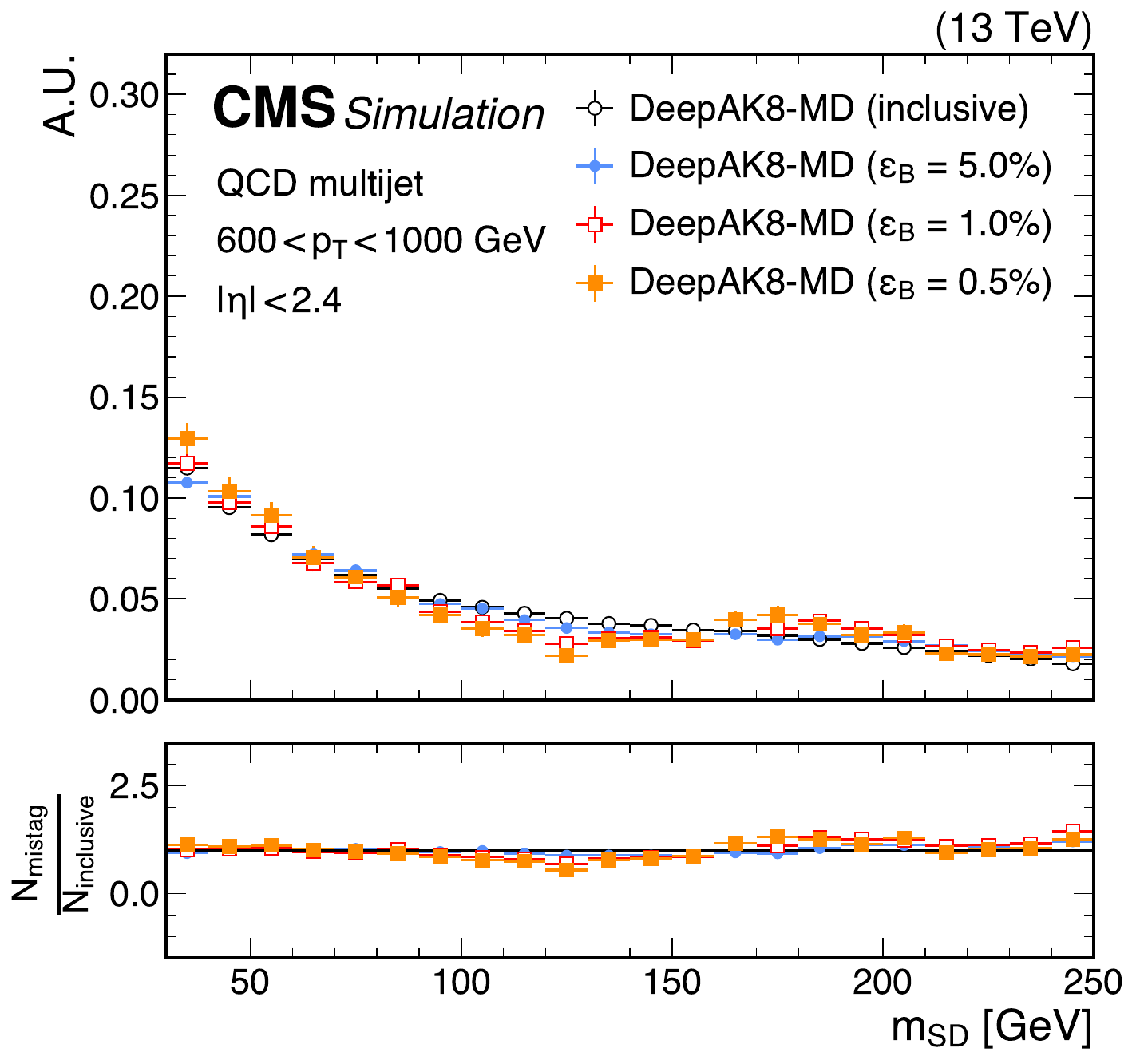}
    \caption{Distributions of \msd for jets from QCD multijet events, in the \pt ranges 200--400\GeV (upper), 400--600\GeV (middle) and 600--1000\GeV (lower), after no selections (``inclusive'') on the \taggername \THWW score (left) and the \deepak score (right) as well as selections corresponding to QCD jet selection efficiencies ($\epsilon_B$) of 5.0\%, 1.0\%, and 0.5\%.
        The error bars represent the statistical uncertainties originating from the limited number of simulated events.
        The lower panels display the ratio of the normalized \msd distributions for the different selection efficiencies ($N_\mathrm{mistag}$) to the normalized inclusive \msd distribution ($N_\mathrm{inclusive}$).
        An offline selection of $\pt > 400\GeV$, $\abs{\eta} < 2.4$, and $\msd > 30\GeV$ is applied to the AK8 jets.
    }
    \label{fig:part_md}
\end{figure}

\begin{figure}[htpb]
    \centering
    \includegraphics[width=0.49\textwidth]{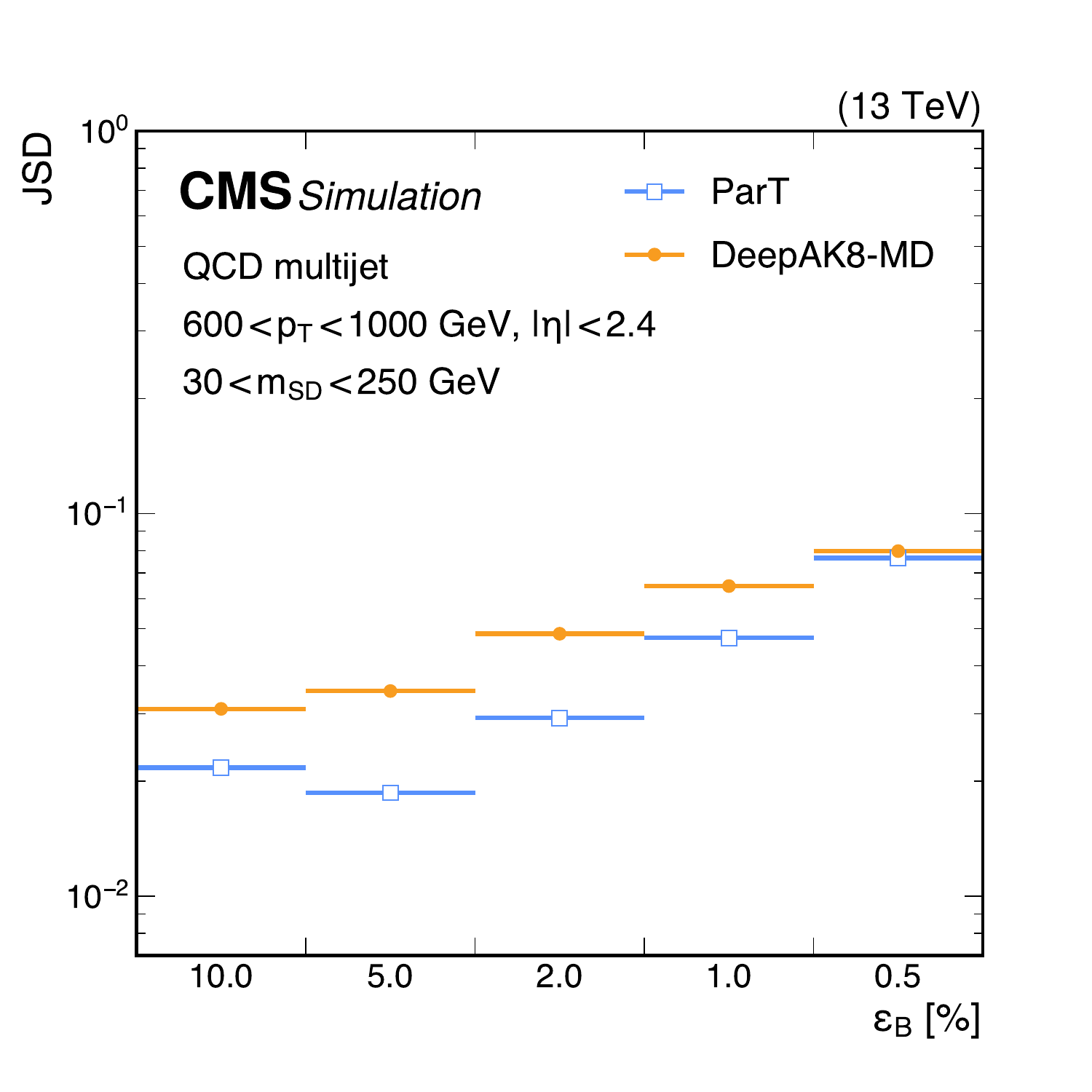}
    \includegraphics[width=0.49\textwidth]{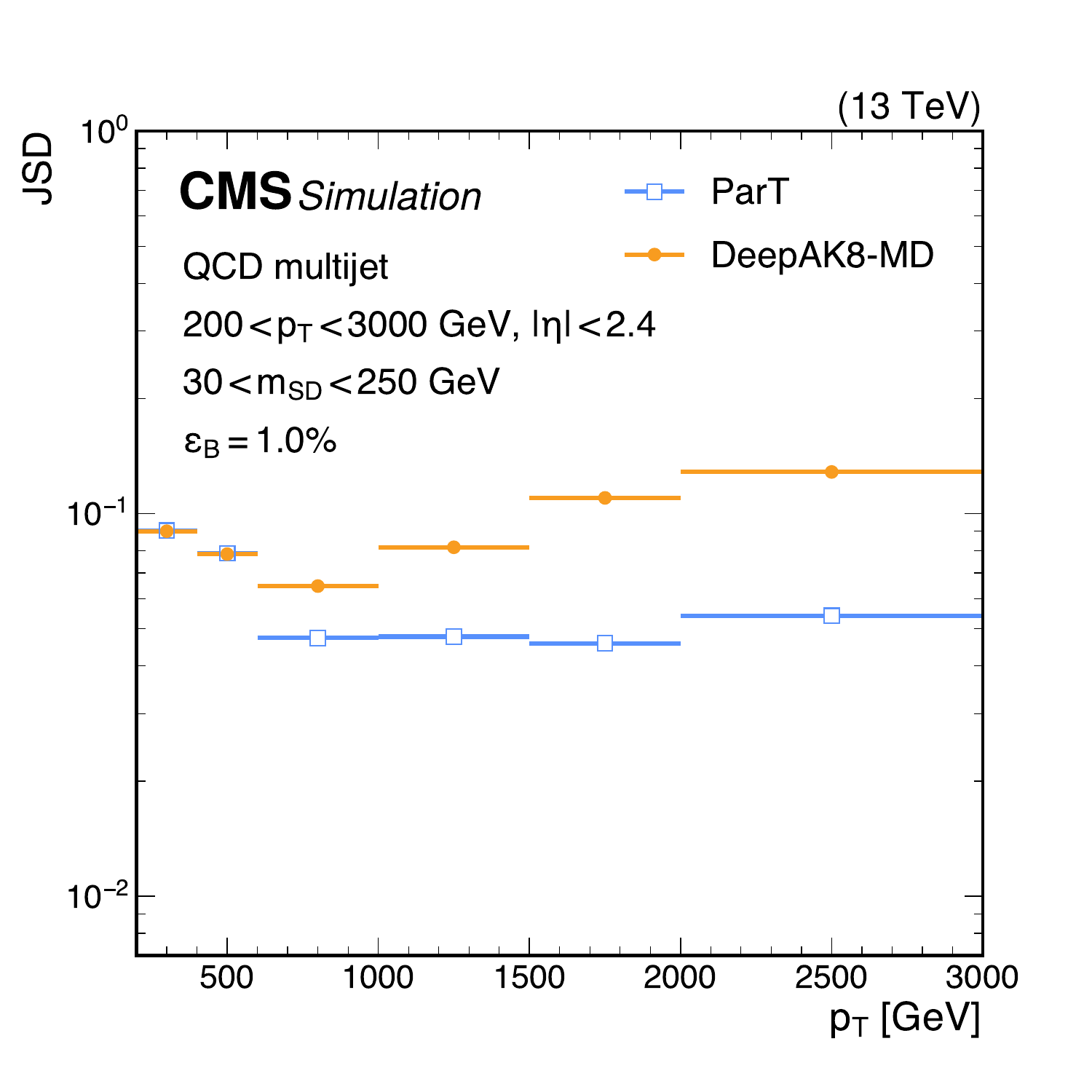}
    \caption{The Jensen--Shannon distance (JSD) using base 2 between the \msd distribution of jets from QCD multijet events with and without a selection on the \taggername and \deepak tagger scores.
        On the left, the JSD is plotted for tagger selections corresponding to different QCD jet selection efficiencies ($\epsb$), with an offline selection of $600 < \pt < 1000\GeV$, $\abs{\eta} < 2.4$, and $30 < \msd < 250\GeV$ applied to the jets.
        On the right, the JSD is plotted for different jet \pt bins, at a fixed \epsb of 1\%.
    }
    \label{fig:part_md_jsd}
\end{figure}

\subsection{Calibration using the Lund jet plane}
\label{sec:part_ljp_method}

Jet identification algorithms are typically calibrated using a similar but more abundant SM process as a proxy for the signal process.
For example, for boosted \hbb calibration, the CMS experiment has used jets produced through \PZ boson decays to \bbbar and through gluon splitting to \bbbar to measure data versus signal simulation disagreement in the tagger score modeling~\cite{CMS:2025kje}.
However, for \hwwq jets, it is difficult to define such a high-purity data sample of four-pronged jets.
Instead, we use a novel method that measures data versus simulation differences in the per-subjet radiation pattern based on densities of their primary LJPs~\cite{Dreyer:2018nbf}, as introduced in Ref.~\cite{CMS:2025eyd}.

    {
        \tolerance=800
        The primary LJP of a jet represents each successive hardest splitting in the 2D ($\ln(1/\Delta)$, $\ln(\kt/\GeVns)$) plane, where $\Delta$ and \kt are the same quantities defined in Section~\ref{sec:tagger_arch}, evaluated here for the emitting and emitted particles of each splitting.
        As discussed in Ref.~\cite{Dreyer:2018nbf}, the primary LJP captures important information about the substructure of the jet.
        The ratio of the densities of primary LJPs of jets from data and simulation are measured in Ref.~\cite{CMS:2025eyd} per subjet in merged two-pronged jets originating from \PW bosons, clustered with the \kt algorithm to two exclusive jets, and binned in subjet \pt.
        \par
    }

\begin{figure}[ht]
    \centering
    \includegraphics[width=\textwidth]{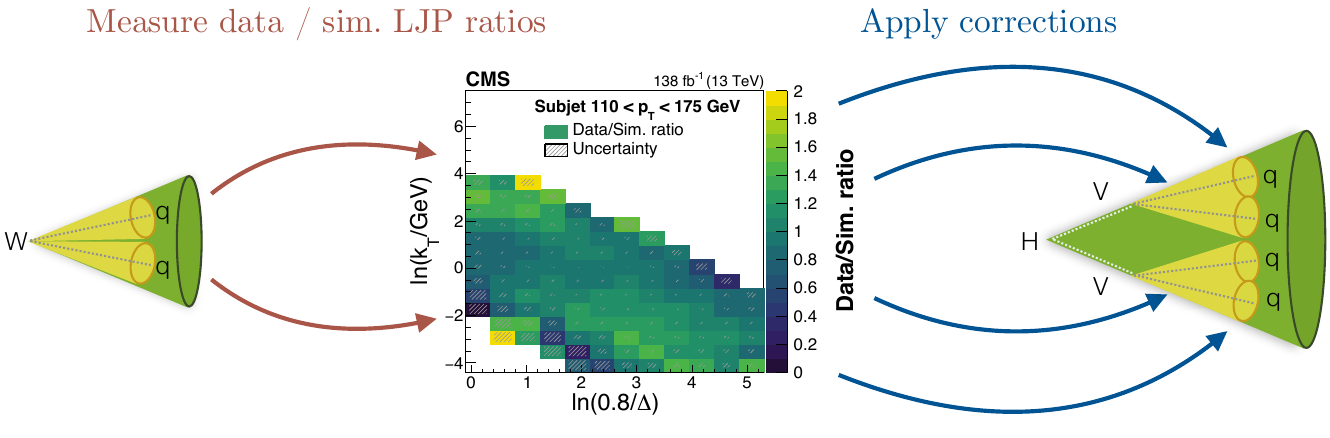}
    \caption{Schematic of the LJP calibration method for \hwws tagging.
        Ratios of primary LJP densities in data and simulation are first measured per subjet in merged two-pronged \PW jets, with an example of such a ratio reproduced from Ref.~\cite{CMS:2025eyd}.
        These are then used to derive correction factors for \hwws signal jets per prong.
    }
    \label{fig:part_ljp_cartoon}
\end{figure}

To measure a correction for the signal efficiency, first, a data-to-simulation relative weight per event is derived for the signal using the primary LJP splittings for each subjet in the \hwws jet, following the procedure described in Ref.~\cite{CMS:2025eyd}.
Each \hwws jet is categorized as either two-, three-, or four-pronged based on the number of quarks within the jet radius, using information from the event generator, and is then reclustered into the corresponding number of subjets.
The relative event weight is defined as the product, taken over all subjets in each splitting, of the data-to-simulation density ratios of primary LJPs, parameterized by \kt, $\Delta$, and the subjet \pt.
Systematic uncertainties related to the validity of this procedure and the quark-subjet correspondence are discussed below.
The final event weights are then either applied to the simulated signal events, as in the top quark validation study shown in Fig.~\ref{fig:part_lundplane} below, or used to derive a single data-to-simulation scale factor (SF), defined as the ratio of the signal yield with and without the event weights, as in the \HH search in Section~\ref{sec:hhbbvv}.
A schematic of the overall calibration procedure is shown in Fig.~\ref{fig:part_ljp_cartoon}.

Uncertainties from the correction procedure are propagated as systematic uncertainties in the SF.
The following sources are considered, as described in detail in Ref.~\cite{CMS:2025eyd}: 1) statistical uncertainties in the derived LJP ratio from the limited number of data events in the \PW boson enriched sample (``ratio stat.''); 2) systematic uncertainties in the ratio because of possible simulation mismodeling (``ratio syst.''), estimated using ``tag and probe'' measurements; 3) uncertainties in the extrapolation of the SF to high \pt subjets (``\pt extrap.''); 4) uncertainties in the true number of prongs inside the \hwws jet (``\# prongs''); 5) whether there exists a correct, one-to-one correspondence between the reclustered subjets and \hwws daughter quarks (``unclust. prongs''); and 6) differences between the LJP densities of the \PW boson jets used for the correction and the \hwws signal jets (``distortion'').
The impacts of these uncertainties for different \HH signals are quantified in Table~\ref{tab:part_ljp_sfs}.
Generally, the uncertainty in the unclustered prongs is dominant, followed by systematic uncertainties in the LJP ratios and the distortion uncertainties.

\begin{table}[!ht]
    \centering
    \caption{Signal efficiency SFs and uncertainties for the BDT selections on the \taggername \hwws tagging outputs in the \HHbbVV search, measured using the LJP calibration method for different \HH signals and analysis regions.
        Both the total combined uncertainty and the components defined in the text are shown.
    }
    \renewcommand{\arraystretch}{1.2}
    \label{tab:part_ljp_sfs}
    \resizebox*{1\textwidth}{!}{
        \begin{tabular}{ccccccccc}
            \multirow{3}{*}{Region} & \multirow{3}{*}{Process} & \multirow{3}{*}{SF}    & \multicolumn{6}{c}{Uncertainty components (fractional)}                                                                                                 \\[0.1mm]
                                    &                          &                        & Ratio                                                   & Ratio  & \pt     & \multirow{2}{*}{\# prongs} & Unclust.             & \multirow{2}{*}{Dist.} \\
                                    &                          &                        & stat.                                                   & syst.  & extrap. &                            & prongs               &                        \\
            \hline
            \multirow{4}{*}{ggF}    & SM ggF \HH               & $0.91^{+0.19}_{-0.13}$ & $0.02$                                                  & $0.08$ & $0.00$  & $0.03$                     & ${}^{+0.17}_{-0.08}$ & ${}^{+0.09}_{-0.08}$   \\
                                    & SM VBF \HH               & $0.94^{+0.22}_{-0.18}$ & $0.02$                                                  & $0.15$ & $0.00$  & $0.01$                     & ${}^{+0.16}_{-0.07}$ & ${}^{+0.07}_{-0.08}$   \\
                                    & VBF \HH ($\kapvv = 0$)   & $0.99^{+0.17}_{-0.08}$ & $0.02$                                                  & $0.02$ & $0.01$  & $0.02$                     & ${}^{+0.17}_{-0.07}$ & ${}^{+0.02}_{-0.00}$   \\
                                    & VBF \HH ($\kapvv = 2$)   & $1.00^{+0.18}_{-0.07}$ & $0.02$                                                  & $0.01$ & $0.01$  & $0.02$                     & ${}^{+0.17}_{-0.07}$ & ${}^{+0.02}_{-0.00}$   \\[3mm]
            \multirow{4}{*}{VBF}    & SM ggF \HH               & $0.83^{+0.22}_{-0.15}$ & $0.03$                                                  & $0.13$ & $0.01$  & $0.04$                     & ${}^{+0.20}_{-0.09}$ & ${}^{+0.09}_{-0.07}$   \\
                                    & SM VBF \HH               & $0.91^{+0.21}_{-0.19}$ & $0.02$                                                  & $0.18$ & $0.00$  & $0.02$                     & ${}^{+0.12}_{-0.05}$ & ${}^{+0.07}_{-0.07}$   \\
                                    & VBF \HH ($\kapvv = 0$)   & $0.88^{+0.14}_{-0.09}$ & $0.02$                                                  & $0.07$ & $0.01$  & $0.02$                     & ${}^{+0.14}_{-0.06}$ & ${}^{+0.03}_{-0.01}$   \\
                                    & VBF \HH ($\kapvv = 2$)   & $0.89^{+0.14}_{-0.08}$ & $0.02$                                                  & $0.06$ & $0.01$  & $0.02$                     & ${}^{+0.13}_{-0.06}$ & ${}^{+0.03}_{-0.01}$\end{tabular}
    }
\end{table}

\subsection{Validation on jets from boosted top quark decays}
\label{sec:part_ljp_validation}

The LJP calibration method is validated for \taggername on jets from decays of boosted top quark candidates.
We define a semileptonic boosted \ttbar enriched data sample, tagging a leptonically decaying top quark ($\PQt\to\PQb\PW\to\PQb\PGm\PGn$), and probing an opposite-side high-\pt AK8 jet representing the hadronically decaying top quark.
The event selection follows that of the control region in Ref.~\cite{CMS:2025eyd}, comprising muon triggers and offline selections for a \PQb-tagged AK4 jet, a leptonically decaying \PW boson---based on the presence of a muon and \ptmiss---and a high-\pt AK8 jet with a mass close to $m_\PQt$.
Jets from simulated \ttbar samples are categorized using generator-level particles as either: ``\PQt-matched''---with all three daughter quarks lying within the jet; ``\PW-matched''---with only the \PW daughter quarks inside the jet; or ``\PQt- and \PW-unmatched''---neither of these two cases.
Only the \PQt-matched jets are reweighted with the LJP ratios, as they are expected to be the most similar to the \hwws signal jets and, indeed, dominate the high-scoring bins in Fig.~\ref{fig:part_lundplane}.

We consider the \THWW discriminant from Eq.~(\ref{eq:thww}), excluding \PTop from the denominator to retain top quark events in the high tagger score bins:
\begin{equation}
    \label{eq:thww_notop}
    \THWWnotop = \frac{\PHWWqqqq + \PHWWqqq}{\PQCD + \PHWWqqqq + \PHWWqqq}.
\end{equation}
The \THWWnotop distributions from the 2017 data sets before and after LJP-reweighting of the events with matched top quark jets are shown in Fig.~\ref{fig:part_lundplane} (left).
The combined uncertainties per bin are also shown in the distributions and data versus simulation ratios.
We observe an overall improvement in the agreement between data and simulation, with the $\chi^2$ test statistic value decreasing from $20.7$ to $16.6$ for 20 degrees of freedom, corresponding to an increase in the two-tailed p-value from $0.415$ to $0.679$.
Importantly, the data and simulation yields are all consistent within statistical and systematic uncertainties in these bins.
We also show the effect of LJP calibration of the \deepak discriminant in Fig.~\ref{fig:part_lundplane} (right), defined similarly as in Eq.~(\ref{eq:deepak}), excluding \PTop from the denominator:
\begin{equation}
    \label{eq:deepak_notop}
    \text{\deepak score (No top)} = \frac{\PWqq + \PHqqqq}{\PQCD + \PWqq + \PHqqqq}.
\end{equation}
We observe a larger improvement in the agreement between data and simulation; however, overall, the agreement is worse than that of \taggername.
The calibration is also performed in the 2016 and 2018 data sets, with similar corrections and improvements in the agreement between data and simulation.

\begin{figure}[ht]
    \centering
    \includegraphics[width=0.49\textwidth]{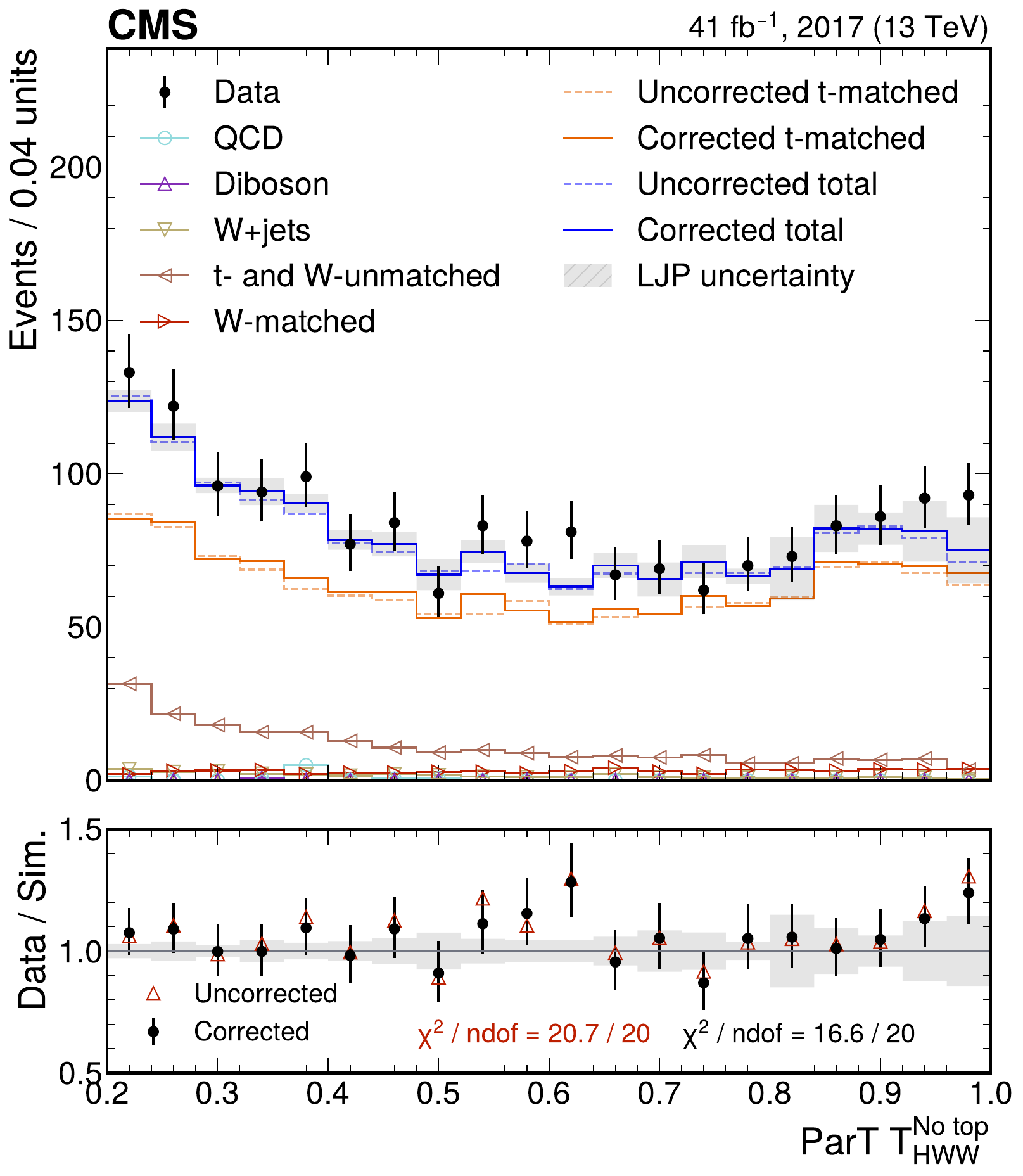}
    \includegraphics[width=0.49\textwidth]{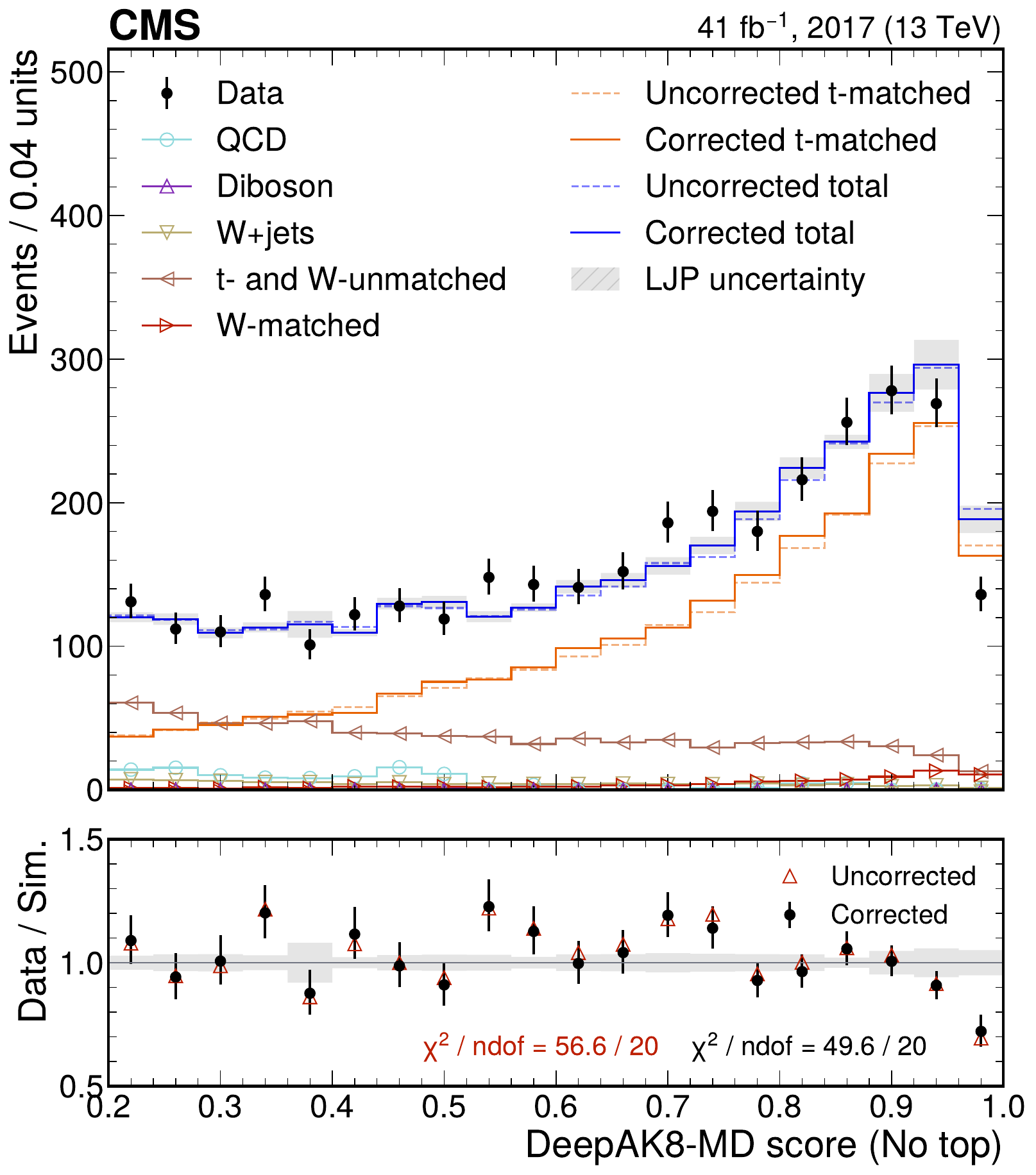}
    \caption{
        Distributions of the \taggername \THWWnotop (left) and \deepak (No top) (right) discriminants with and without the LJP corrections for \PQt-matched jets for data and individual simulated processes in the upper panels, and data versus simulation ratios in the lower panels.
        The combined uncertainties from LJP-based SFs per bin are shown in shaded gray, and the statistical uncertainty in the number of data events per bin is represented by vertical error bars in the upper and lower panels.
        The $\chi^2$ test statistic values between data and simulation, normalized to the number of degrees of freedom (ndof), are also shown for both discriminants with and without LJP corrections.
    }
    \label{fig:part_lundplane}
\end{figure}

\section{Impact on measuring \texorpdfstring{\HHbbVV}{HH to bbWW*} production properties}
\label{sec:hhbbvv}

Targeting boosted \hwws production is motivated by a variety of BSM scenarios, including the production of heavy resonance decays of the form $\PX \to \PH\PY$, where \PX and \PY are new scalar resonances with $\PH\to\bbbar$ and $\PY\to\PV\PV\to4\Pq$~\cite{CMS:2026shw}, and measurements of Higgs boson couplings through the production of Higgs boson pairs~\cite{CMS:2025ngq}.
One property of the Higgs boson that is particularly challenging to measure is its quartic \HHVV coupling.
Deviations to this coupling can significantly enhance VBF \HH, particularly at high \HH invariant mass, which is why recent searches for boosted $\HH\to\bbbb$ have been successful in constraining this coupling~\cite{CMS:2022gjd,ATLAS-CONF-2024-003}.
This motivates further exploration of boosted \HH production, particularly for the first time in the all-hadronic \bbww channel~\cite{CMS:2025ngq}, which represents the largest branching fraction of \HH decays after \bbbb.
However, such a search necessitates powerful classification algorithms for both \hbb and \wws to discriminate against the dominant QCD multijet background.
This section summarizes the use of \taggername in the first all-hadronic \HHbbVV search; a more detailed description of the search can be found in Section~4.4.2 of Ref.~\cite{CMS:2025ngq}.

The primary physics objects considered in the search are two large-radius (AK8) jets, each representing one of the two Higgs bosons, which are tagged as \hbb candidates and \hwws candidates using DL algorithms.
Offline selections are defined using a selection on the \hbb tagging score of the \bbbar candidate jet, and multivariate selections using a boosted decision tree (BDT) on the \taggername scores of the \wws candidate, along with kinematic variables, targeting the ggF and VBF production modes.
Working points for the BDT are optimized for expected limits for ggF and VBF \HH cross sections, with signal (background) efficiencies of 7\% (${\approx}0.01\%$) and 20\% (${\approx}0.003\%$), respectively.
As the \taggername scores of the \wws candidate rank highest in feature importance among all the inputs to the BDT, the \taggername algorithm is critical to achieving this level of signal-to-background discrimination.

Several sources of theoretical and experimental systematic uncertainties are considered in the signal and background modeling in the signal regions.
Overall, the dominant uncertainty in the \HH signal strength is the statistical uncertainty in the QCD multijet background estimation, driven by the limited data sample size in the signal region, with an impact on the best fit signal strength of 50\% relative to the overall uncertainty.
The next largest source of uncertainty is in the signal efficiency of the BDT selections on the \taggername \hwws tagging outputs, measured using the LJP method.
A summary of the calculated SFs and uncertainties for different \HH signals and signal regions is provided in Table~\ref{tab:part_ljp_sfs}, where the overall uncertainty ranges between 15 and 25\%, and has a relative impact of 23\% on the best fit signal strength.
Among all the channels included in the CMS \HH combination, the all-hadronic \bbww channel provides the second most stringent constraint on the \kapvv coupling of $[-0.04, 2.05]$ at 68\% confidence level (\CL), after the boosted \bbbb channel ($[0.62, 1.41]$ at 68\% \CL)~\cite{CMS:2025ngq}.

\section{Summary}
\label{sec:conclusion}

The particle transformer (\taggername) deep neural network for classifying a wide variety of jets from decays of Lorentz-boosted resonances has been presented.
In particular, \taggername enables effective identification of all-hadronic Higgs boson to \PW boson (\hwwq) decays by the CMS experiment for the first time.
A novel training strategy is used to address challenges pertaining to \hwws classification, through which \taggername achieves $>$50\% \hwwq selection efficiency with a multijet background efficiency of 1\%, while maintaining decorrelation with the jet mass.
The performance is calibrated on data using the primary Lund jet planes of individual subjets, with data-to-simulation scale factors measured in the 0.9--1.0 range, and relative uncertainties between 7 and 23\%.
The \taggername algorithm represents a significant advancement in the identification capabilities of multiprong jets from highly boosted resonances in CMS, illustrated by the first search for boosted Higgs boson pair production in the all-hadronic \bbww channel.

\begin{acknowledgments}
We congratulate our colleagues in the CERN accelerator departments for the excellent performance of the LHC and thank the technical and administrative staffs at CERN and at other CMS institutes for their contributions to the success of the CMS effort. In addition, we gratefully acknowledge the computing centers and personnel of the Worldwide LHC Computing Grid and other centers for delivering so effectively the computing infrastructure essential to our analyses. Finally, we acknowledge the enduring support for the construction and operation of the LHC, the CMS detector, and the supporting computing infrastructure provided by the following funding agencies: SC (Armenia), BMBWF and FWF (Austria); FNRS and FWO (Belgium); CNPq, CAPES, FAPERJ, FAPERGS, and FAPESP (Brazil); MES and BNSF (Bulgaria); CERN; CAS, MoST, and NSFC (China); MINCIENCIAS (Colombia); MSES and CSF (Croatia); RIF (Cyprus); SENESCYT (Ecuador); ERC PRG and PSG, TARISTU24-TK10 and MoER TK202 (Estonia); Academy of Finland, MEC, and HIP (Finland); CEA and CNRS/IN2P3 (France); SRNSF (Georgia); BMFTR, DFG, and HGF (Germany); GSRI (Greece); MATE and NKFIH (Hungary); DAE and DST (India); IPM (Iran); SFI (Ireland); INFN (Italy); MSIT and NRF (Republic of Korea); MES (Latvia); LMTLT (Lithuania); MOE and UM (Malaysia); BUAP, CINVESTAV, CONACYT, LNS, SEP, and UASLP-FAI (Mexico); MOS (Montenegro); MBIE (New Zealand); PAEC (Pakistan); MSHE, NSC, and NAWA (Poland); FCT (Portugal); MESTD (Serbia); MICIU/AEI and PCTI (Spain); MOSTR (Sri Lanka); Swiss Funding Agencies (Switzerland); MST (Taipei); MHESI (Thailand); TUBITAK and TENMAK (T\"{u}rkiye); NASU (Ukraine); STFC (United Kingdom); DOE and NSF (USA).

\begin{sloppypar}
\setlength\emergencystretch{\hsize}
\hyphenation{Rachada-pisek} Individuals have received support from the Marie-Curie program and the European Research Council and Horizon 2020 Grant, contract Nos.\ 675440, 724704, 752730, 758316, 765710, 824093, 101115353, 101002207, 101001205, and COST Action CA16108 (European Union); the Leventis Foundation; the Alfred P.\ Sloan Foundation; the Alexander von Humboldt Foundation; the Science Committee, project no. 22rl-037 (Armenia); the Fonds pour la Formation \`a la Recherche dans l'Industrie et dans l'Agriculture (FRIA) and Fonds voor Wetenschappelijk Onderzoek contract No. 1228724N (Belgium); the Beijing Municipal Science \& Technology Commission, No. Z191100007219010, the Fundamental Research Funds for the Central Universities, the Ministry of Science and Technology of China under Grant No. 2023YFA1605804, the Natural Science Foundation of China under Grant No. 12535004, and USTC Research Funds of the Double First-Class Initiative No.\ YD2030002017 (China); the Ministry of Education, Youth and Sports (MEYS) of the Czech Republic; the Shota Rustaveli National Science Foundation, grant FR-22-985 (Georgia); the Deutsche Forschungsgemeinschaft (DFG), among others, under Germany's Excellence Strategy -- EXC 2121 ``Quantum Universe" -- 390833306, and under project number 400140256 - GRK2497; the Hellenic Foundation for Research and Innovation (HFRI), Project Number 2288 (Greece); the Hungarian Academy of Sciences, the New National Excellence Program - \'UNKP, the NKFIH research grants K 131991, K 138136, K 143460, K 143477, K 147557, K 146913, K 146914, K 147048, TKP2021-NKTA-64, and 2025-1.1.5-NEMZ\_KI-2025-00004, and MATE KKP and KKPCs Research Excellence and Flagship Research Groups grants (Hungary); the Council of Science and Industrial Research, India; ICSC -- National Research Center for High Performance Computing, Big Data and Quantum Computing, FAIR -- Future Artificial Intelligence Research, and CUP I53D23001070006 (Mission 4 Component 1), funded by the NextGenerationEU program, the Italian Ministry of University and Research (MUR) under Bando PRIN 2022 -- CUP I53C24002390006, PRIN PRIMULA 2022RBYK7T (Italy); the Latvian Council of Science; the Ministry of Science and Higher Education, project no. 2022/WK/14, and the National Science Center, contracts Opus 2021/41/B/ST2/01369, 2021/43/B/ST2/01552, 2023/49/B/ST2/03273, and the NAWA contract BPN/PPO/2021/1/00011 (Poland); the Funda\c{c}\~ao para a Ci\^encia e a Tecnologia (Portugal); the National Priorities Research Program by Qatar National Research Fund; MICIU/AEI/10.13039/501100011033, ERDF/EU, ``European Union NextGenerationEU/PRTR", projects PID2022-142604OB-C21, PID2022-139519OB-C21, PID2023-147706NB-I00, PID2023-148896NB-I00, PID2023-146983NB-I00, PID2023-147115NB-I00, PID2023-148418NB-C41, PID2023-148418NB-C42, PID2023-148418NB-C43, PID2023-148418NB-C44, PID2024-158190NB-C22, RYC2021-033305-I, RYC2024-048719-I, CNS2023-144781, CNS2024-154769 and Plan de Ciencia, Tecnolog{\'i}a e Innovaci{\'o}n de Asturias, Spain; the Chulalongkorn Academic into Its 2nd Century Project Advancement Project, the National Science, Research and Innovation Fund program IND\_FF\_68\_369\_2300\_097, and the Program Management Unit for Human Resources \& Institutional Development, Research and Innovation, grant B39G680009 (Thailand); the Eric \& Wendy Schmidt Fund for Strategic Innovation through the CERN Next Generation Triggers project under grant agreement number SIF-2023-004; the Kavli Foundation; the Nvidia Corporation; the SuperMicro Corporation; the Welch Foundation, contract C-1845; and the Weston Havens Foundation (USA).
\end{sloppypar}
\end{acknowledgments}

\bibliography{auto_generated}
\cleardoublepage \appendix\section{The CMS Collaboration \label{app:collab}}\begin{sloppypar}\hyphenpenalty=5000\widowpenalty=500\clubpenalty=5000\cmsinstitute{Yerevan Physics Institute, Yerevan, Armenia}
{\tolerance=6000
A.~Hayrapetyan, V.~Makarenko\cmsorcid{0000-0002-8406-8605}, A.~Tumasyan\cmsAuthorMark{1}\cmsorcid{0009-0000-0684-6742}
\par}
\cmsinstitute{Institut f\"{u}r Hochenergiephysik, Vienna, Austria}
{\tolerance=6000
W.~Adam\cmsorcid{0000-0001-9099-4341}, L.~Benato\cmsorcid{0000-0001-5135-7489}, T.~Bergauer\cmsorcid{0000-0002-5786-0293}, M.~Dragicevic\cmsorcid{0000-0003-1967-6783}, C.~Giordano\cmsorcid{0000-0001-6317-2481}, P.S.~Hussain\cmsorcid{0000-0002-4825-5278}, M.~Jeitler\cmsAuthorMark{2}\cmsorcid{0000-0002-5141-9560}, N.~Krammer\cmsorcid{0000-0002-0548-0985}, A.~Li\cmsorcid{0000-0002-4547-116X}, D.~Liko\cmsorcid{0000-0002-3380-473X}, M.~Matthewman, J.~Schieck\cmsAuthorMark{2}\cmsorcid{0000-0002-1058-8093}, R.~Sch\"{o}fbeck\cmsAuthorMark{2}\cmsorcid{0000-0002-2332-8784}, M.~Shooshtari\cmsorcid{0009-0004-8882-4887}, M.~Sonawane\cmsorcid{0000-0003-0510-7010}, W.~Waltenberger\cmsorcid{0000-0002-6215-7228}, C.E.~Wulz\cmsAuthorMark{2}\cmsorcid{0000-0001-9226-5812}
% ...
\par}
$^{1}$Also at Yerevan State University, Yerevan, Armenia\\
$^{2}$Also at Technische Universit\"{a}t Wien, Vienna, Austria\\
% ...
\end{sloppypar}
\end{document}